\documentclass[alpha-refs]{wiley-article}
\usepackage{hyperref}
\hypersetup{colorlinks=false,pdfborder={0 0 0}}

\usepackage[utf8]{inputenc}
\usepackage[english]{babel}

\usepackage{siunitx}
\usepackage[export]{adjustbox}
\usepackage{amsmath}

\newcommand\myabstract[1]{
\ifdefined\linenoyes
\begin{internallinenumbers}[1]
#1
\end{internallinenumbers}
\fi
\unless\ifdefined\linenoyes
#1
\fi
}

\paperfield{Quarterly Journal of the Royal Meteorological Society}
\corraddress{Leif Denby, Institute for Climate and Atmospheric Science, School of Earth and Environment, University of Leeds, Leeds, LS2 9JT, United Kingdom}
\corremail{l.c.denby@leeds.ac.uk}
\fundinginfo{Met Office/NERC Grant NE/N013840/1}

\papertype{Research Article}

\title{Characterising the shape, size and orientation of cloud-feeding coherent boundary layer structures}

\author[1]{Leif Denby}
\author[1]{Steven J. B{\"o}ing}
\author[1]{Douglas J. Parker}
\author[1]{Andrew N. Ross}
\author[2]{Steven M. Tobias}

\affil[1]{School of Earth and Environment, University of Leeds}
\affil[2]{School of Mathematics, University of Leeds}

\runningauthor{Leif Denby}

\begin{document}

\maketitle
\selectlanguage{english}
\begin{abstract}

\myabstract{
This paper presents two techniques for characterisation of cloud-feeding
coherent boundary layer structures through analysis of large-eddy simulations of
shallow cumulus clouds, contrasting conditions with and without ambient shear.
The first technique is a generalisation of the two-point correlation function where
the correlation length-scale as well as orientation can be extracted.
The second technique decomposes the vertical transport by coherent structures
by the shape, size and orientation of these structures.
It is found that the structures dominating the vertical flux are plume-like in
character (extending from the surface into cloud), show small width/thickness asymmetry
and rise near-vertically in the absence of ambient wind.
The planar stretching and tilting of boundary layer structures
caused by the introduction of ambient shear is also quantified, demonstrating the
general applicability of the techniques for future study of other boundary
layer patterns.

}

\end{abstract}%

\ifdefined\linenoyes
\linenumbers
\fi

\section{Introduction}\label{introduction}

Coherent boundary layer structures carry perturbations of temperature,
moisture and vertical velocity necessary to trigger convective clouds, by
overcoming the boundary layer top inversion in a conditionally unstable
atmosphere. However, the degree to which the spatial distribution, morphology
and perturbations carried by the coherent structures affect how clouds form is
currently uncertain, as is which external drivers affect these properties of
the coherent structures. To study comprehensively the formation of clouds from
coherent boundary layer structures, we must first be able to identify and
measure the properties of these structures, which is the aim of this paper.

Coherent structures in the boundary layer carry so-called non-local 
(cannot simply be calculated from local scalar values) counter-gradient 
transport in the boundary layer \citep{Deardorff1966}, providing transport against the vertical 
mean gradient of moisture and heat (in contrast to smaller turbulent eddies 
doing down-gradient, diffusive, transport). 
Owing to the limited resolution available in Global Circulation Models 
and Numerical Weather Prediction models, it is necessary to parameterise 
the unresolved sub-grid processes that provide vertical transport and 
lead to convective cloud formation.
The development of parametrisations of non-local transport has been key to
improving boundary layer 
parametrisations \citep{Holtslag1991, Brown1997}. 
Over the past two decades, the Eddy-Diffusivity Mass-Flux (EDMF) approach 
to boundary layer parametrisation \citep[e.g.][]{Siebesma2007,Rio2008,Neggers2009a,Neggers2009b,Rio2010} has become
popular: in this approach, local turbulent transport and transport 
by coherent structures (leading to the formation 
of convective clouds) are modelled separately. However, our current lack of 
knowledge of how these coherent structures are affected by external forcings
(ambient wind, surface heterogeneity, cold-pools, etc) and changes in these
structures affect cloud formation, limits our ability 
to refine models of the non-local transport, and thus in representing the 
genesis (formation) of convective clouds in weather and climate simulations.

As well as influencing the formation of individual clouds, these structures
capture the convective state of the atmosphere, through their spatial
organisation and by persisting sub-grid length-scales of motion over time (a form of convective "memory").
Representation of these sub-grid forms of organisation are largely
absent in contemporary convection parameterisations, however the importance of
convective organisation in affecting, for example, the radiative properties of
the atmosphere, and the impact of limited representation of these processes
in models, is becoming increasingly clear \citep{Bony2017}.

Prior work on characterising the morphology of coherent boundary layer structures
has focused primarily on measuring coherence in the boundary layer as a
whole, not looking at the properties of individual coherent structures, but instead
producing bulk length-scale estimates using spectral peaks in the autocorrelation 
and covariance spectrum to measure spatial and angular coherence (in the horizontal plane). 
\citet{Jonker1999} found in cloud-free Large-Eddy Simulations (LES) that the shortest correlation length-scale exists in vertical velocity on the order of the
boundary layer depth, whereas the
buoyancy providing field (potential temperature $\theta$ for dry LES, virtual
potential temperature $\theta_v$ when water vapour is included) typically
attains larger steady-state length-scales \citep{Jonker1999,
DeRoode2004, Salesky2017}.
\Citet{DeRoode2004} in addition found that for passive tracers in dry and
stratocumulus topped boundary layers the vertical profile of horizontal length-scales is determined by
the surface to boundary layer top buoyancy flux ratio
$r=\frac{\overline{w'\theta_v'}_T}{\overline{w'\theta_v'}_0}$, with minimum
length-scales attained when this ratio was $r\approx-0.2$ for dry convection
(which is the classical buoyancy flux-ratio scale found for dry convective
boundary layers) and $r\approx-1$ in the case of stratocumulus.

In simulations where clouds are present the picture becomes more complicated
for the dynamic scalars \citep{SchumannMoeng1991, DeRoode2004}, apart
from vertical velocity which stays largely unchanged (restricted dynamically by
the boundary layer depth).
The buoyancy ($\theta_v$) no longer necessarily
reaches a steady-state length-scale profile, and water-vapour and potential
temperature produce meso-scale variability (which appear to cancel in their
contributions to $\theta_v$).
The length-scale characteristics found in LES have been supported through
similar analysis of observations \citep{NichollsLemone1980, LenschowSun2007}, 
with the relative scales of $q$ and $\theta$ changing
with Bowen ratio, producing narrower scales for the scalar dominating
buoyancy. Identifying and characterising coherent structures in observations,
\citet{SchumannMoeng1991} found that coherence length-scale typically peaks mid
boundary layer, where the number of structures reaches a minimum.
\citet{Miao2006} found the plume spacing and width to be $~0.7h_{BL}$ and
$0.2h_{BL}$ respectively mid boundary layer.

In place of studying correlation in inverse distance (or wave-length) space,
the present work studies correlation in real space using cumulants \citep[see history and review by][]{Lauritzen2007}.
Cumulants have been applied by \citet{Lohou2000} to study thermals in a daytime 
boundary layer with weak wind shear over land, and characterise the influence of
anisotropy on vertical transport. \citet{Schmidt1989} used cumulants to study 
coherence in the vertical rather than the horizontal plane, and identified both 
large-scale plumes and transient thermals in a convective boundary layer. 
They have also been used to expand the prognostic equations \citep{Ait-Chaalal2016a}
to predict correlations between fields rather than the fields themselves.
In modelling by \citet{Lohou2000} and \citet{Schmidt1989} correlation is studied in 
real space, rather than inverse distance, or wave-length space. In this work 
we develop this technique further by utilising cumulants and producing vertical 
profiles of integral length-scale and orientation of coherence in real space.

With respect to identifying individual coherent structures in the boundary layer
prior work focused on using limit values on vertical velocity 
or water vapour concentration (either separately or in combination) to define object masks
\citep{BergStull2004, GrantBrown1999, NichollsLemone1980, SchumannMoeng1991}.
More recently \citet{Efstathiou2020} developed a masking technique to maximise the
vertical transport carried by the selected region of the boundary layer. 
\citet{Couvreux2010} noted that object masks based on the physical fields (vertical 
velocity, water vapour, temperature) had the drawback that they poorly capture 
transport through the boundary layer top inversion and into the cloud-layer, and 
proposed a technique based on a surface-released decaying passive tracer to 
track the rising boundary layer structures.
This tracer technique has been used to identify coherent structures 
in cloudy boundary layers \citep{Dawe2012, Park2016, Brient2019}. 
The current work uses the tracer in combination with the object-splitting technique 
described in \citet{Park2018}, to identify individual structures that are characterised 
by a prominent local maximum of the vertical velocity in the boundary layer. 

The aim of this paper is to demonstrate the use of new techniques to
characterise the morphology of coherent structures in the boundary layer and 
thereby provide the means to identify the properties of structures 
which dominate the vertical transport.
We first demonstrate that using measures of spatial coherence on individual
fields gives an incomplete picture of the properties of coherent boundary
layer structures, next move to characterise individual coherent structures
and later decompose the vertical transport by the structure characteristics.
Having the ability to measure the shape, size and orientation of the
coherent structures which dominate the vertical transport will enable the
study in further work of how and to what extent these properties effect how clouds form and
organise, and how external factors effect the coherent structures. This in
turn will enable the construction of clearer process-understanding for the
coupling of clouds to the boundary layer, which ultimately can be used to build
better representations of clouds in weather and climate models.

As a means of investigating the extent to which the methods discussed
herein are able to unpick and quantify boundary layer transport and its
influence on clouds in different environmental conditions, we will use
two simulations of shallow convection, with and without shear, as a demonstration
of large-scale influence on coherent boundary layer structures and
convective clouds. The modelling setup for the simulations are discussed
in Sec. \ref{sec:modelling-setup}. The methods used to identify coherent
boundary layer structures and quantify their properties will be
discussed in Sec. \ref{sec:methods}. Finally the application of these
methods will be shown in Sec. \ref{sec:results} and a discussion of this
analysis will be given in Sec. \ref{sec:discussion}.

\section{Modelling setup}\label{modelling-setup}

\label{sec:modelling-setup}

Simulations were carried out with the non-hydrostatic UCLA-LES
Large-Eddy Simulations model \citep{Stevens2005} with
two-moment warm-rain microphysics scheme \citep{Stevens2008} on a
$\SI{40}{km} \times \SI{40}{km} \times \SI{4}{km}$ double-periodic
domain with an isotropic grid-spacing of $(\Delta x, \Delta y, \Delta z)=
(25, 25, 25) \SI{}{m}$.

\begin{figure}
\centering
\includegraphics[width=0.7\textwidth]{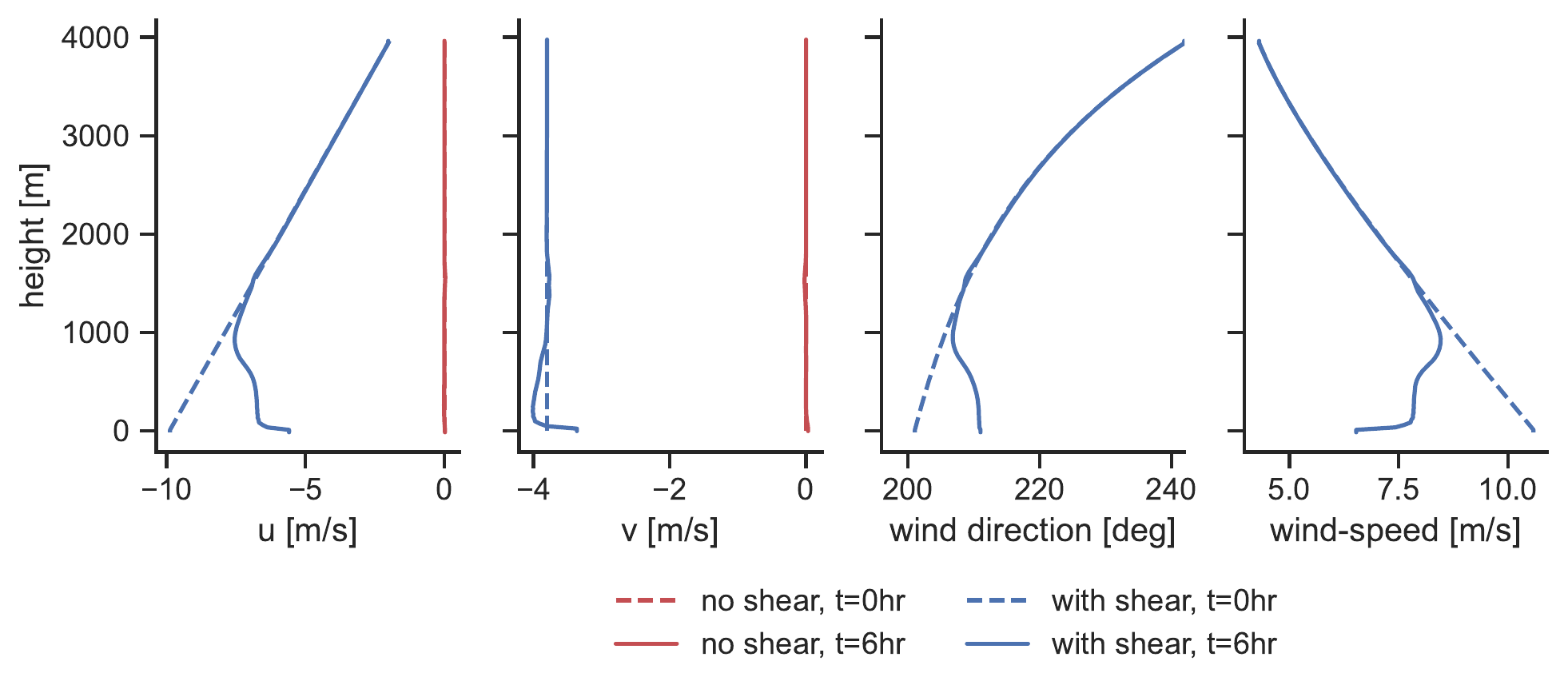}
\caption{\label{fig:wind-profiles} Vertical profiles of horizontal mean
wind at times $t = \SI{0}{\hour}$ and $t = \SI{6}{\hour}$ in meridional ($v$) 
and zonal ($u$) directions, together with wind direction $\phi$ (measured from $x$-axis),
for cases with and without shear.
Mean wind direction in sub-cloud layer after $t = \SI{6}{\hour}$ 
is at $\phi \approx \SI{210}{\degree}$ in simulation with shear.}
\end{figure}

The simulation setup is based on the Rain in Cumulus over the Ocean (RICO) 
field study \citep{Rauber2007a} and associated LES model inter-comparison 
study \citep{VanZanten2011}.
The RICO setup is characterised by shallow cumuli developing from moisture-dominated
fluxes from the ocean surface, with the clouds constrained in
growth by a prescribed large-scale subsidence aloft and large-scale
advection of moisture out of the domain. In the original
inter-comparison study, the simulation settles into a quasi-steady state
after a short ($\approx \SI{2}{\hour}$) rapid response to the initial
condition, after which convection slowly (over $\approx \SI{20}{\hour}$)
aggregates into larger cloud clusters by precipitation-induced cold pools. As
the process leading to formation of these cloud clusters is not the focus 
of this study, we will be considering only the stage of cloud development
before these large clusters have developed (here using $t = \SI{6}{\hour}$).

In this work there are two key differences to the original RICO setup as 
published in \citet{Seifert2015}.
Firstly, in order to study the effect of ambient wind-shear on the
coherent boundary layer structures, two simulations were run, one with and
one without shear (see profiles in Fig. \ref{fig:wind-profiles}). In the
former the wind-profile from RICO was left unchanged and in the latter
the meridional and zonal wind components were set to zero. Secondly,
because the near-surface horizontal velocity differs between the two
simulations, the bulk aerodynamic parameterisation of surface flux was
replaced with a fixed sensible ($F_s=\SI{7}{W/m^2}$) and latent heat flux
($F_v=\SI{150}{W/m^2}$) so that the two conditions have the same
fluxes provided from the surface. The surface flux values were estimated
from the original RICO simulation once near-equilibrium conditions have
been reached (at $t \approx \SI{6}{\hour}$).

\begin{figure}
\centering
\includegraphics[width=0.59\textwidth]{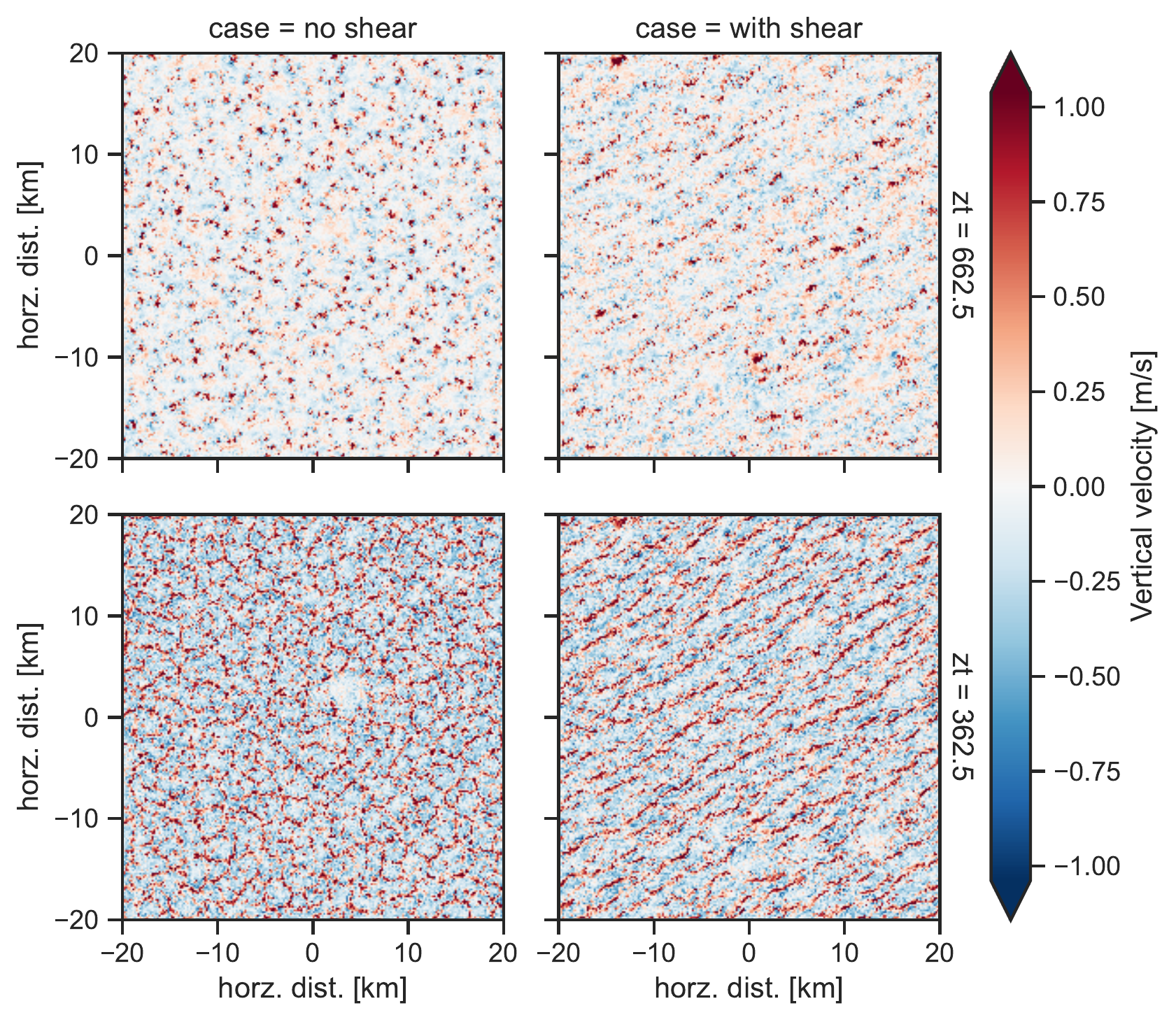}
\caption{\label{fig:overview-cross-sections} Horizontal cross-sections
of vertical velocity through the boundary layer middle
($z\approx \SI{362.5}{\meter}$, bottom) and at cloud-base height
($z \approx \SI{662.5}{\meter}$, top) for simulations with shear (right) and
without (left). The presence of ambient shear is seen to clearly break
the geometry of convective cells and create elongated coherent boundary
structures.}
\end{figure}

As seen in the horizontal cross-sections of
vertical velocity in Fig. \ref{fig:overview-cross-sections}, the presence
of ambient shear causes both boundary layer structures and clouds to become
organised into elongated structures instead of convective cells. This is
noticeable through the elongated line-like regions of high vertical velocity in
simulations with shear, causing the clouds (cloud-base is at
$z\approx \SI{650}{\meter}$) to organise into structures resembling \emph{cloud streets} 
instead of (as in the case without shear) at the nodes of boundary layer convective
cells. The development of cloud streets under conditions with ambient shear
is consistent with prior studies \citep[see reviews by][]{Etling1993, Young2002},
as is the presence of sheet-like coherent 
structures attached to the surface extending into the bulk of the 
boundary layer \citep[as in][]{Khanna1998}.

\section{Methods}\label{methods}

\label{sec:methods}

Before the properties of coherent boundary layer structures that
trigger clouds can be measured, it is necessary to define exactly what
we mean by a coherent structure. In contrast to the cloud layer, where we can define a coherent structure purely on the concentration of water droplets, in the boundary layer there are at least three scalar fields that carry the perturbation that eventually triggers a cloud: moisture, temperature and vertical velocity.

We first investigate the length-scales of variability in these fields as
a bulk-property of the fluid. If these fields were to vary over similar
length-scales it would be relatively straightforward to define threshold
criteria on either of the scalar fields that would apply to all
fields. Unfortunately 
this is not the case (section \ref{sec:characteristic-length-scales}) and so we instead develop a method which identifies air with properties similar to that which triggers
convective clouds (section \ref{sec:object-identification}), by tracking
air entering newly-formed convective clouds.

\subsection{Characteristic length-scales - cumulant analysis}
\label{sec:characteristic-length-scales}

As an alternative to moments, cumulants provide a means to summarise the
statistical correlation between one or more variables \citep{Lauritzen2007}.
Similarly to \citet{Tobias2016}, where cumulants were used to identify and 
measure coherent structures in 3D rotating Couette flow, we here utilise the second 
cumulant (two-point correlation function), which for fields $\psi$ and $\varphi$ at 
height $z$ \citep[here $z_1=z_2=z$ in contrast to][]{Tobias2016} is given by
\begin{equation}
\begin{split}
c_{\psi\varphi}(\xi, \nu, z) = &\\
\frac{1}{L_x L_y} &\int_{0}^{L_x} \int_{0}^{L_y} \psi'(x,y,z) \varphi'(x + \xi, y + \nu, z) dx dy \nonumber,
\end{split}
\end{equation}
where $\psi'$ and $\varphi'$ are deviations from the horizontal mean of $\psi$
and $\varphi$ respectively, and $L_x$ and $L_y$ are the lengths of the domain in
the $x$- and $y$-direction. The positions are wrapped around in the $x$- 
and $y$-direction exploiting the periodic boundary conditions of the simulation.

\begin{figure}
\centering
\includegraphics[width=0.7\textwidth]{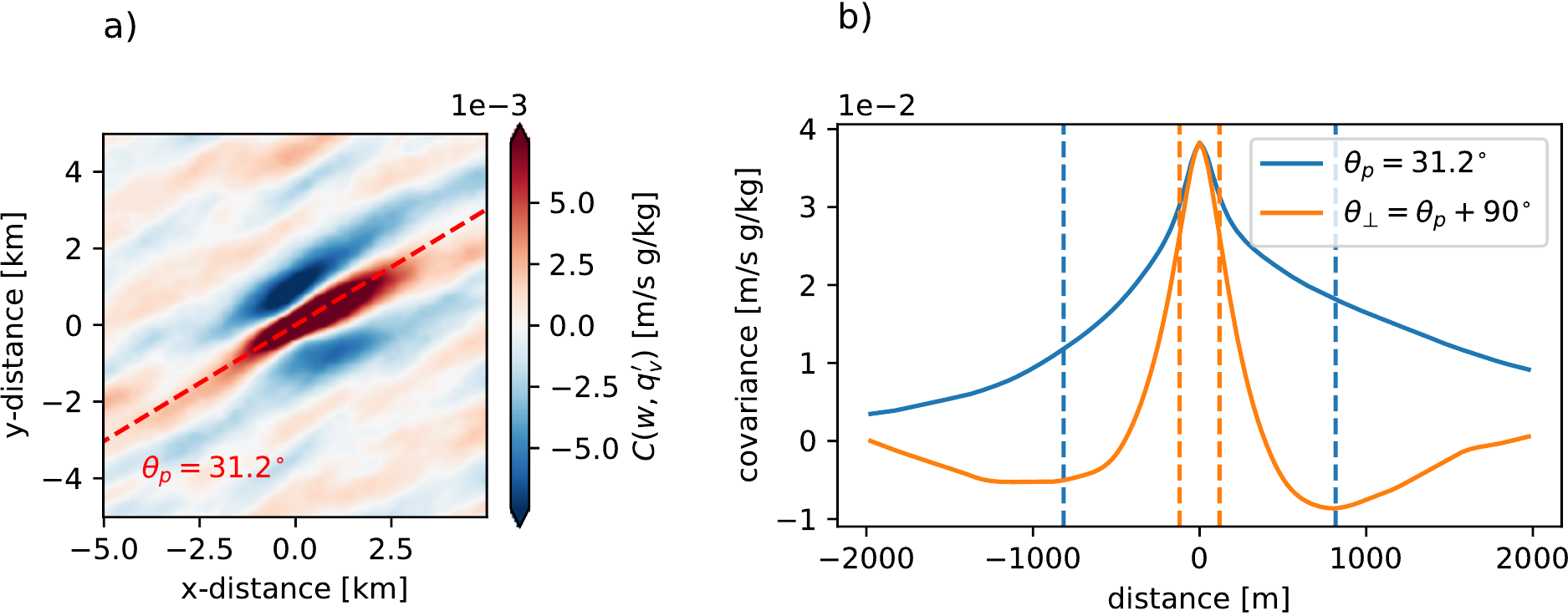}
\caption{\label{fig:cumulant-example} a) Cumulant of vertical velocity and water vapour
(i.e. the horizontal moisture flux) in horizontal plane at $z = \SI{300}{\meter}$ with principal
direction of coherence identified by the red dashed line, and b) the same cumulant
sampled along (in blue) and perpendicular (in orange) to the principal axis
with characteristic (integral) width indicated with vertical lines. 
The elongated nature of the coherence as seen on the left is
quantified by a significantly larger ($\approx\SI{1500}{m}$ vs
$\approx\SI{500}{m}$) characteristic length along the direction of shear. In both a) and b)
the scaling of correlation magnitude is written above.}
\end{figure}

An example of this method applied to the spatial correlation of vertical
velocity ($\psi=w$) and water vapour ($\varphi=q_v$) in the middle of the
boundary layer in a simulation with ambient shear is shown in Fig.
\ref{fig:cumulant-example}. In cases such as this where an external forcing is
causing boundary layer and cloud structures to develop in a preferential
direction, the cumulant will show increased correlation in this direction.
To quantify this asymmetry we identify a \emph{principal} and
\emph{perpendicular} direction of coherence (measured in terms of the angles
$\theta_p$ and $\theta_\perp$) of the
central part of the cumulant. This central part $\hat{c}_{\psi\varphi}$ is defined
as the connected region at the origin with the same sign as at the origin of the
cumulant $c_{\psi\varphi}$. Treating $\hat{c}_{\psi\varphi}$ as a
2D mass-distribution, we then estimate the orientation angle as the principal
axis (eigenvector with largest eigenvalue) of the moment of inertia tensor:

\begin{equation}
\overline{\overline{I}} = \begin{bmatrix}
\int \hat{c}_{\psi\varphi}(\xi,\nu)\nu^2\ d\xi d\nu & \int \hat{c}_{\psi\varphi}(\xi,\nu)\xi\nu\ d\xi d\nu \\
\int \hat{c}_{\psi\varphi}(\xi,\nu)\xi\nu\ d\xi d\nu & \int \hat{c}_{\psi\varphi}(\xi,\nu)\xi^2\ d\xi d\nu
\end{bmatrix}.
\end{equation}

The cumulant can then be sampled in this horizontal plane along the principal 
and perpendicular directions of coherence
(as seen in Fig. \ref{fig:cumulant-example} right) so that
the coherence can be quantified in these directions. The presence of
ambient shear is for example seen to cause elongation in the direction
of the ambient wind (this will be discussed in detail in section \ref{sec:results-characteristic-length-scales}).

Once the direction of principal coherence ($\theta_p$) has been identified, a
characteristic length-scale ($L_p$) of coherence may be estimated in
this direction and similarly in the perpendicular direction ($L_\bot$).
These length-scales are computed through a cumulant-weighted integral
of distance ($l = \sqrt{\xi^2 + \nu^2}$) from the cumulant origin:

\begin{align}
L^{\psi,\varphi}_\delta &= \frac{\int_{-\frac{L}{2}}^{\frac{L}{2}} l\ \hat{c}^{\delta}_{\psi,\varphi}(l)\ dl}{\int_{-\frac{L}{2}}^{\frac{L}{2}} \hat{c}^{\delta}_{\psi,\varphi}(l)\ dl},
\end{align}
where $\delta \in [p, \bot]$ (for either the principal or perpendicular 
length-scale) and $L=min(L_x, L_y)$ (the minimum of the simulation domain length in the $x$- and $y$-direction), with the cumulant along a particular
direction given by
\begin{align}
\hat{c}^{\delta}_{\psi,\varphi}(l) &= \hat{c}_{\psi,\varphi}(\xi{=}l\cos(\theta_\delta), \nu{=}l\sin(\theta_\delta))
\end{align}
evaluated at an arbitrary point using piece-wise linear interpolation. To measure the
degree of elongation we define the \textit{asymmetry ratio} $r_a = \frac{L_p}{L_\bot}$.

As well as providing a means to quantify the length-scale of correlation
(and thus a single characteristic length-scale for all coherent boundary
layer structures), the shape of the cumulant can provide insight into the
dynamical structure of coherent structures by quantifying the relative spatial
distribution of different scalar fields. This is possible by studying
two different aspects of the cumulant produced from two different
fields, specifically the offset of the cumulant peak value from the
origin and the skewness of the distribution around the origin. An offset
of the cumulant from the origin indicates that the extreme values of two
different scalar fields are located spatially offset from each other and
would suggest something is driving a separation between two fields.
Similarly, skewness in the cumulant distribution indicates that the two
different scalar fields have differently spatially skewed distributions,
e.g one field may appear spatially Gaussian, but another may be skewed
relative to this. The corollary to this is that a 2\textsuperscript{nd} 
cumulant, between
two different scalar fields, which is centered on and symmetric around
the origin, indicates that these two fields on average are spatially
distributed identically (in terms of skewness) around their respective 
peak values and their peak values are, on average, co-located.

\subsection{Object-based analysis}
\label{sec:object-based-analysis}

To gain a more comprehensive understanding of transport by coherent structures
in the boundary layer, we transition from looking at the boundary layer air in
a bulk sense, to studying transport by individual coherent structures that
may trigger clouds.
This requires identifying the regions of the boundary layer that contribute to
transport into convective clouds, splitting these regions into individual
coherent structures and finally formulating methods to quantify the properties
of these structures.

\subsubsection{Object identification}\label{object-identification}

\label{sec:object-identification}

To quantify the characteristic properties of individual coherent structures
carrying out vertical transport, these structures must first be identified. This
was achieved by first producing a 3D mask to pick out the part of the atmosphere
thought to contain coherent structures, and thereafter splitting this mask into
individual 3D objects.

The 3D mask was produced from the concentration of a passive tracer ($\phi$)
decaying with a time-scale $\tau$, which was released from the surface \citep[as 
first used in][]{Couvreux2010}. Specifically the time evolution of the tracer is given by
\begin{equation}
\frac{\partial \phi}{\partial t} = -\frac{\phi}{\tau}.
\end{equation}
The decay time-scale was set to $\tau = \SI{15}{min}$ in this study as this
represents the typical overturning time-scale of boundary layer eddies in
the simulations used (see appendix \ref{appendix:convective-timescale}).

From the scalar $\phi$ a mask is created using its standard deviation in a
horizontal cross-section ($\sigma_\phi(z)$) and its local deviation
from the horizontal mean
($\phi'(x,y,z) = \phi(x,y,z) - \overline{\phi(z)}$) by requiring that
the local deviation is $n$ standard deviations from the mean, i.e.~the
mask $m(x,y,z)$ is given by
\begin{equation}
m(x,y,z) =
\left\{
\begin{matrix}
1 & \text{if} & \phi'(x,y,z) > n\ \sigma_\phi(x,y,z), \\
0 & \text{otherwise}
\end{matrix}
\right.
\end{equation}
here $n=2$ was used as this was found to produce closest agreement between
the properties of air entering clouds and those identified to belong to
coherent structures (see section \ref{section:cross-correlation-of-scalar-fields}).
The choice of decay time-scale and limit value for $n$ is similar to those 
($\tau = \SI{15}{min}$ and $n=2.5$)
identified by \citet{Chinita2018} to be optimal when studying shallow 
moist convection in the BOMEX (Barbados Oceanographic and Meteorological 
Experiment) case \citep{Siebesma2003}.

The constructed 3D mask was observed to identify boundary layer air with
thermodynamic properties similar to air entering into recently formed clouds (see
section \ref{sec:identifying-cloud-feeding-structures} for details),
making it a suitable method to separate vertical transport by local
diffusive mixing (small eddies) from transport by larger, non-local eddies carrying
fluxes leading to cloud-formation.

The method of \citet{Park2018} was used to identify individual objects from the 3D mask.
This method works by first labelling contiguous regions of the mask as \textit{proto-objects}.
These proto-objects are then further subdivided based on a second scalar field (here
vertical velocity $w$) by assigning points to local maxima in the second scalar and
splitting where a boundary (the "col") between two maxima has a relative value below a
predefined threshold \citep[$f=0.7$ was used for this "col"-factor, as in][]{Park2018}.

To measure the thermodynamic properties of air causing the formation of clouds,
individual clouds were identified from 2D column-integrated liquid water path
($m_{lwp}$) with a threshold value of $m_{lwp} > \SI{0.01}{kg/m^2}$
and tracked by spatial overlap in consecutive time-steps using the
method detailed in \citet{Heus2013}. This method identifies
\emph{active} clouds as ones with at least one buoyant core (identified
from the virtual potential temperature $\theta_v$) and \emph{passive} clouds
as those without a buoyant core.
In addition, this cloud-tracking method splits clouds
with multiple buoyant cores into smaller sub-clouds with the non-buoyant
regions defined as \emph{outflow}. In this work we only consider the
properties of air entering single-core \emph{active} clouds
\citep[reaching the level of free convection, see][]{Stull1985}
as these are
likely to have the strongest and clearest connection to boundary layer
variations. To facilitate selecting clouds which recently formed we also keep
track of the age of each cloud ($t_{age}$) by storing the time of appearance
for each tracked cloud.


\subsubsection{Object characterisation}
\label{sec:object-characterisation}

Once individual 3D coherent structures have been identified, a method is needed to
calculate characteristic properties of these objects. Here we detail
techniques to compute characteristic length-scales and orientation 
of each coherent structure.

\paragraph{\textbf{Topological measures - Minkowski functionals}}
\label{sec:methods-minkowski-functionals}
Instead of attempting to fit a parameterised shape (for example an
ellipsoid) to each object with the intention of estimating an object's
scale (length, width and thickness), we instead calculate a set of
characteristic scales using the so-called \emph{Minkowski functionals}
\citep{Minkowski1903} that measure the topology of arbitrary structures in
$N$-dimensional space (see review by \citet{Mecke2000} for details).  
These have been used in other physical applications to characterise for example
dissipative structures in magnetohydrodynamic turbulence \citep{Zhdankin2014a}, 
galaxy distribution \citep{Schmalzing1997} and cosmological structure 
formation \citep{Sahni1998, Schmalzing1999}.
In three dimensions the Minkowski functionals are
\begin{align}
\label{eqn:mink_0}
V_0 &= V = \int dV, \\
\label{eqn:mink_1}
V_1 &= \frac{A}{6} = \frac{1}{6} \int dS, \\
\label{eqn:mink_2}
V_2 &= \frac{H}{3 \pi} = - \frac{1}{6 \pi} \int \nabla \cdot \hat{\mathbf{n}} dS = \frac{1}{6\pi}\int (\kappa_1 + \kappa_2)dS, \\
\label{eqn:mink_3}
V_3 &=  \frac{1}{4 \pi} \int (\kappa_1 \kappa_2) dS = \chi,
\end{align}
where $\hat{\mathbf{n}}$ is the surface normal, $\kappa_1$, $\kappa_2$ are the maximum and minimum local curvature, and $\chi$ the Euler characteristic (essentially related to the number of
holes through an object).
To evaluate these integrals numerically on the discrete output from the large-eddy
simulations, we use Crofton's formula (see Appendix \ref{appendix:croftons-formula}),
which provides discrete approximations for terms (for example the surface 
normal) which are otherwise difficult to evaluate on objects constructed from
individual cubic volumes of the simulation underlying grid. For reference,
the Minkowski functionals for a parameterised spheroid and ellipsoid will be shown, using
the analytical expressions (where available, and otherwise numerical integration)
for surface area and mean curvature given in \citet{Schmalzing1999} and \citet{Poelaert2011}.

From these functionals a characteristic \emph{length} ($L_m$), \emph{width}
($W_m$) and \emph{thickness} ($T_m$) can be calculated as
\begin{equation}
L_m = \frac{3 V_2}{4 V_3}, W_m = \frac{2 V_1}{\pi V_2}, T_m = \frac{V_0}{2V_1},
\end{equation}
where the normalisation is such that all of these measures equal the radius when
applied to a sphere.  Note that for objects with one or more holes $V_3 = \chi \le 0$ and so the length will not be evaluated for these objects.
The length, width and thickness may be further summarised by computing 
the \emph{filamentarity} ($F_m$) and \emph{planarity} ($P_m$)
\begin{equation}
F_m = \frac{L_m - W_m}{L_m + W_m}, P_m = \frac{W_m - T_m}{W_m + T_m},
\end{equation}
which in turn indicate whether an object is more \emph{stick}-like
(large filamentarity) or \emph{pancake}-like (large planarity).
The Minkowski functionals thus enable the quantification of an object's
shape, making it possible to, for example, investigate whether the objects
that are the primary contributors to vertical transport have a
characteristic shape, which can be used to inform the formulation of an
integral model for representing this transport.

\begin{figure}
\centering
\includegraphics[width=0.95\textwidth]{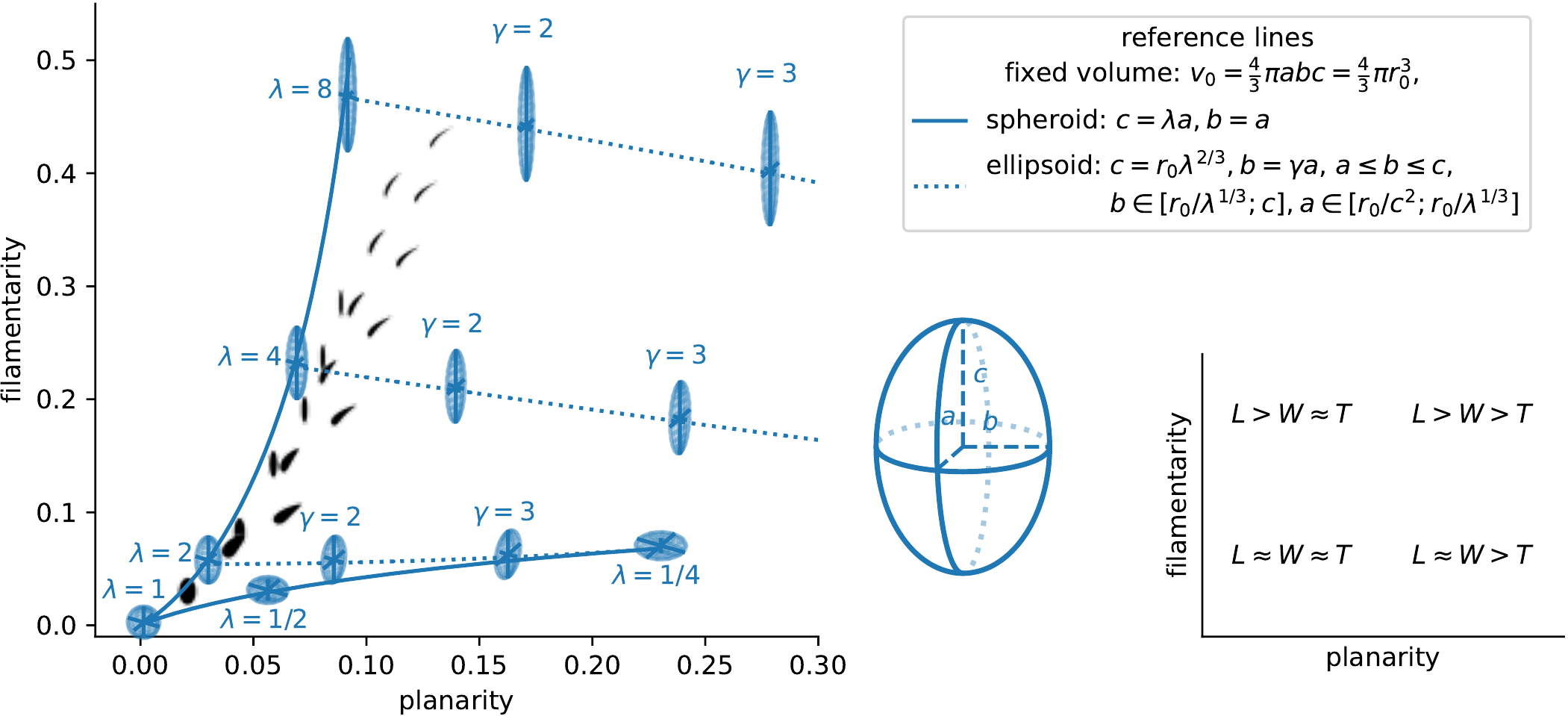}
\caption{\label{fig:fp-synthetic-structures} Filamentarity vs planarity
plot for synthetically created 3D thermal structures with centerline
length $\SI{500}{\meter}$ and varying shear and distance (left). For each structure the
data point is marked with a vertical cross-section (black). Two sets of reference
lines are given using analytical solutions to the Minkowski functionals of a 
spheroid (solid blue) and ellipsoid (dotted blue). $\lambda$ parameterises the aspect
ratio of the spheroid semi- and major axis. $\gamma$ parameterisises the deformation
of a spheroid with a fixed value $\lambda$ and fixed major axis, by varying the
remaining two axis (with aspect ratio $\gamma$). The filamentarity-planarity values
effectively enable the projection of an object's length ($L$), thickness ($T$) and width ($W$)
into two coordinates (right).}
\end{figure}

To provide a reference for the range of values which may be expected for
coherent structures in the atmospheric boundary layer, Fig.
\ref{fig:fp-synthetic-structures} is a filamentarity vs planarity plot
for a number of synthetically created, numerically integrated, sheared and
stretched spheroids. The length (through the center of the structure)
was kept constant across all
shapes while varying the width and shearing distance. Each data point is
marked by an outline of the shape's structure by plotting the vertical
cross-section through each shape's symmetry plane. As a means of
reference for the numerically integrated shapes the analytical functions
for filamentarity and planarity are plotted for a spheroid, by
integrating the analytical forms of the Minkowski functionals for a
spheroid while varying the aspect ratio between one axis and the two
others. In addition to the spheroid reference lines, the deformation of
a spheroid through a general ellipsoid while keeping the volume and major
axis length constant is provided, including the aspect ratio of the
two remaining axis as $\alpha$. It can for example be seen that a
prolate spheroid with $\lambda=2$ becomes an oblate spheroid with
$\lambda=1/4$ (the Minkowski functionals are independent of orientation,
which will be dicussed next).

\paragraph{\textbf{Object tilt and orientation}}

\begin{figure}
\centering
\includegraphics[width=0.7\textwidth]{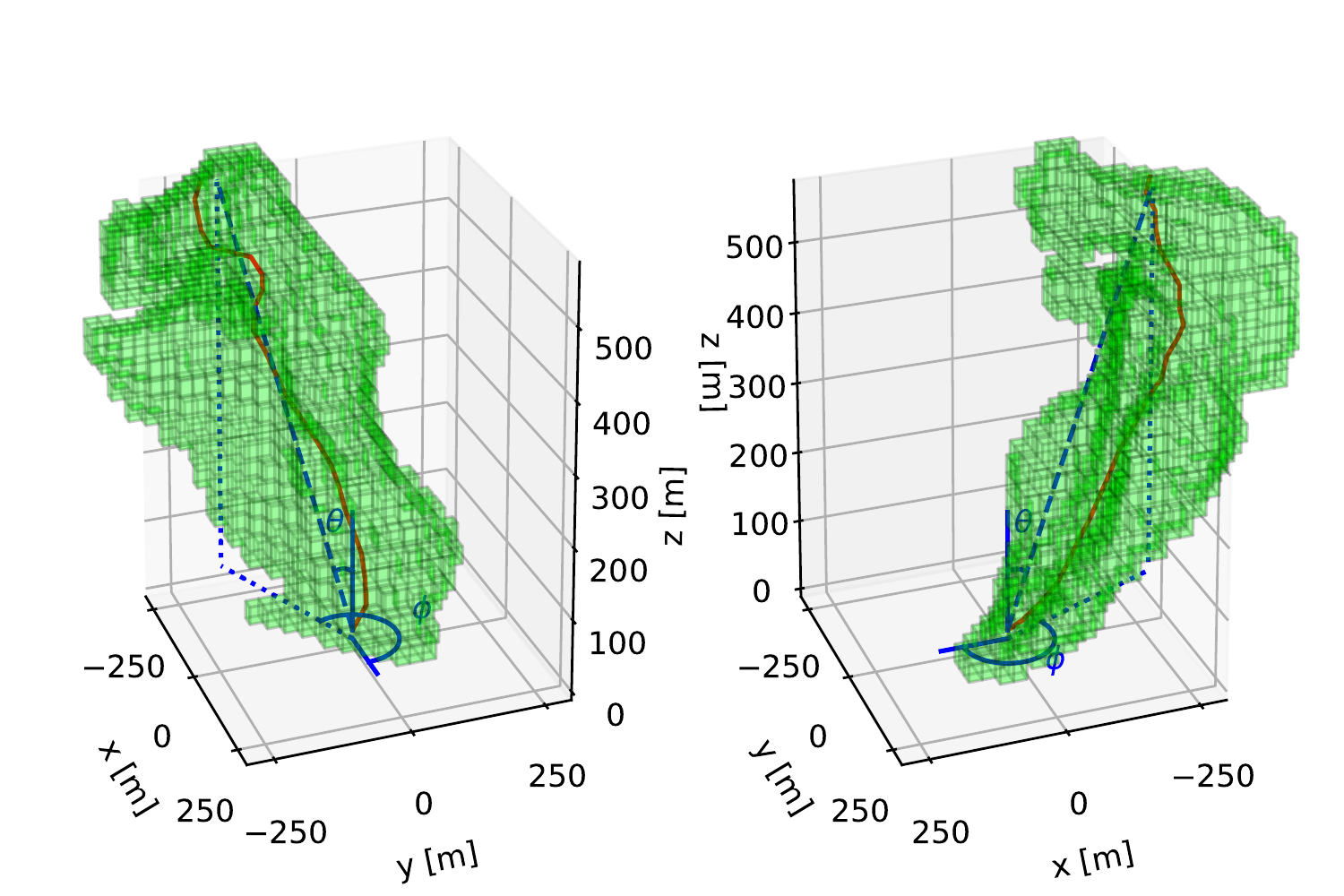}
\caption{\label{fig:object_orientation_calc_example}
Voxel-rendering of single coherent structure from sheared simulation from
two viewing angles, with object
orientation angles $\theta$ (tilt from $z$-axis) and $\phi$ (xy-orientation from $x$-axis)
and centerline (red) indicated.
}
\end{figure}
As the Minkowski functionals (detailed in the previous section) only provide
measures of scale, but not orientation, of individual objects, we introduce
a means of calculating the $xy$-plane orientation ($\phi$, measured from the
$x$-axis) and tilt ($\theta$, measured from the $z$-axis) of an individual object
(see \autoref{fig:object_orientation_calc_example}). 
The characteristic tilt and orientation for an individual coherent structure
is calculated by first forming a \emph{centerline} through the centroids
of vertically adjacent slices of a given structure, and then computing the
area-weighted angular average of $z$-tilt and $xy$-orientation along this
centreline (see appendix \ref{appendix:slope-and-orientation-calc} for
details).

\section{Results}\label{results}

\label{sec:results}

In this section we analyse the sheared and non-sheared simulations using
the methods detailed above. 
The analysis first focuses on extracting characteristic length-scales of
different fields in the boundary layer as a whole, without attempting to
identify individual coherent structures.  Later the boundary layer air that is
likely to be feeding convective clouds is identified.  And lastly, properties of
\emph{individual} coherent structures are studied with the aim of revealing the
properties of coherent structures that dominate the vertical moisture flux.
We focus here on the moisture flux as the simulations represent convection over
the ocean and so have a low Bowen ratio. Similar analysis may be carried
out for heat flux in conditions of high Bowen ratio.

\subsection{Vertical profiles of characteristic horizontal scales}

\label{sec:results-characteristic-length-scales}

\begin{figure*}
\centering
\includegraphics[width=\textwidth]{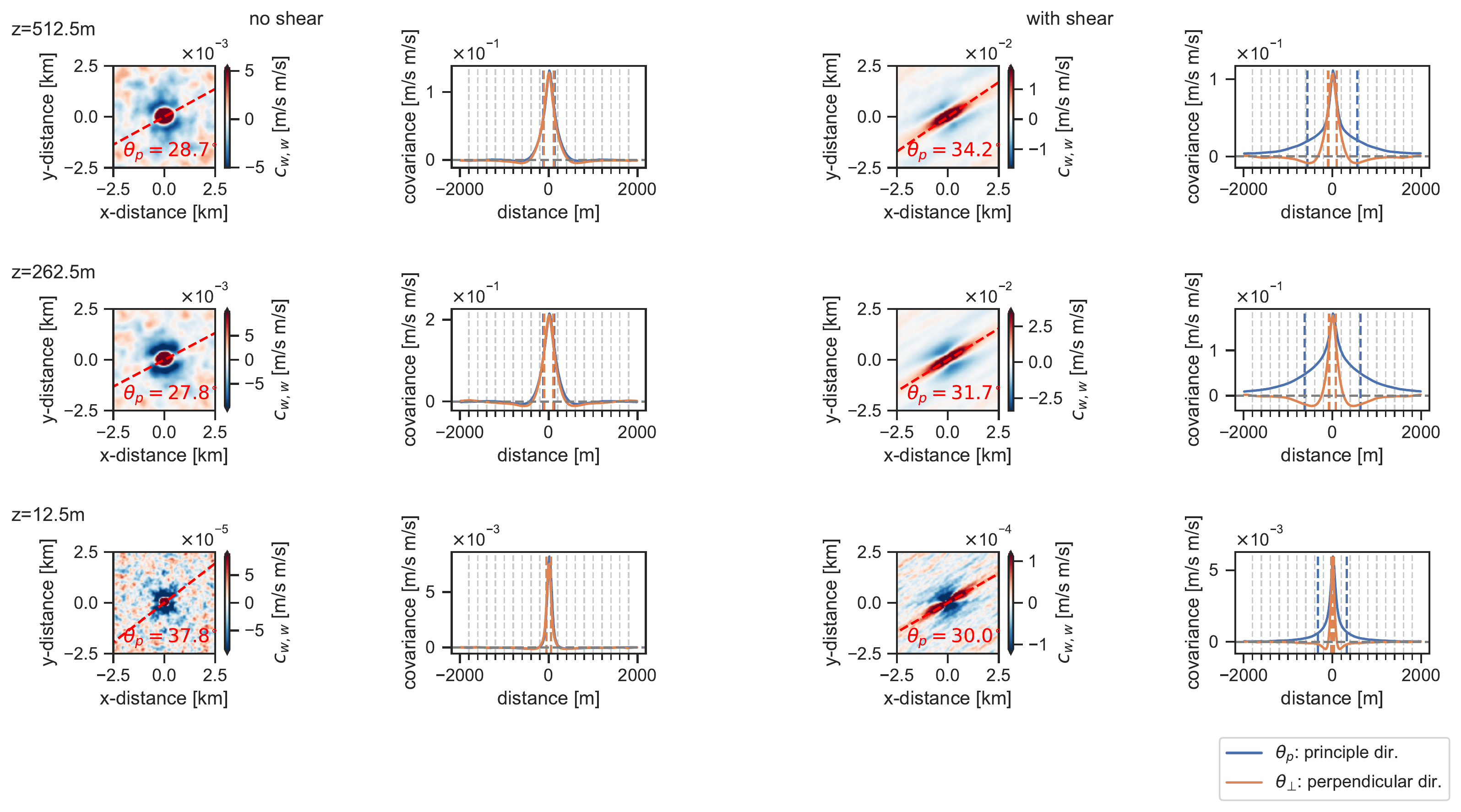}
\caption{\label{fig:cumulant-cross-sections-vertical-velocity}
2\textsuperscript{nd} cumulant of vertical velocity with vertical velocity at increasing heights in the boundary layer in simulations without shear (leftmost two columns) and with shear (rightmost two columns). At each height the cumulant (left, measuring coherence as
a function of distance) is associated with the same cumulant sampled along the identified
principle and perpendicular direction of coherence (right), with the calculated coherence
length-scale indicated with slashed vertical lines. Note the magnitude of coherence changes
with height. For each subfigure the scaling of the correlation magnitude is written above.
}
\end{figure*}

\begin{figure*}
\centering
\includegraphics[width=\textwidth]{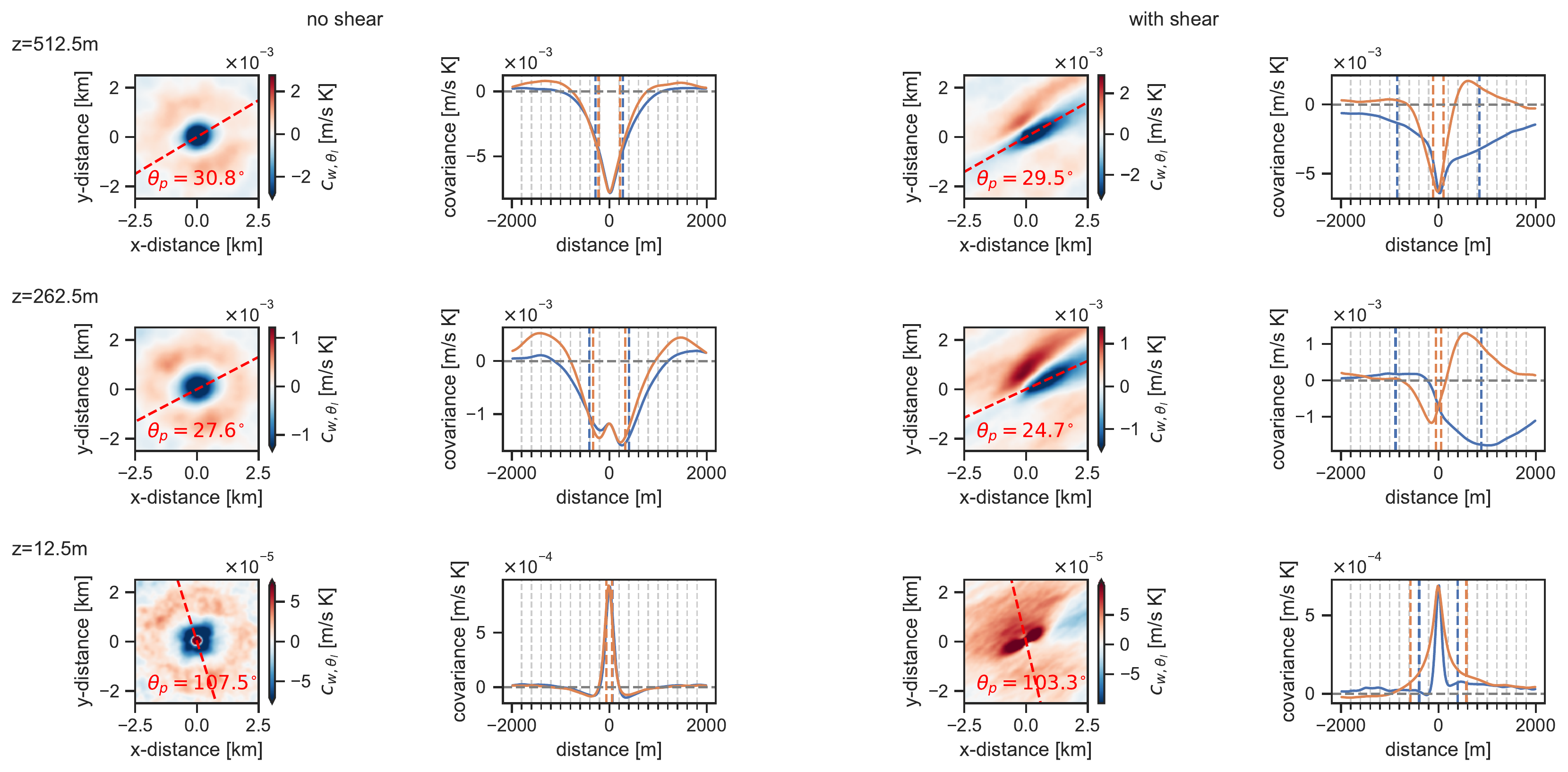}
\caption{\label{fig:cumulant-cross-sections-potential-temperature}
2\textsuperscript{nd} cumulant of vertical velocity with liquid water potential temperature (panels as in Figure \ref{fig:cumulant-cross-sections-vertical-velocity})
}
\end{figure*}

To study the vertical transport produced by coherent structures in the
boundary layer, we must first formulate how to define these structures.
Unfortunately the different scalar fields that are connected with
transport relevant to moist convection (vertical velocity, buoyancy,
moisture and heat content), vary on very different length-scales and these
scales change with height in the boundary layer. To demonstrate this
variability, Figure
\ref{fig:cumulant-cross-sections-vertical-velocity} and
\ref{fig:cumulant-cross-sections-potential-temperature} show the 2\textsuperscript{nd}
cumulant of vertical velocity with itself ($c_{w,w}$, the auto-correlation)
and the 2\textsuperscript{nd} cumulant of vertical velocity and liquid 
water potential temperature ($c_{w,\theta_l}$) in horizontal plane mid, 
boundary layer ($z=\SI{300}{\meter}$).

Considering first the auto-correlation of vertical velocity ($c_{w,w}$), we
note that vertical velocity features are elongated with ambient shear and
axisymmetric without shear, as expected.
The coherence length-scale (slashed vertical line) is largest in the
middle of the boundary layer where vertical velocity peaks, before thermals are
decelerated becoming negatively buoyant in the relatively warm and dry 
layer below cloud.

Considering instead $c_{w,\theta_l}$, there is narrow length-scale of positive 
correlation until $z\approx
\SI{200}{\meter}$, embedded within a larger-scale negative correlation.  The positive
correlation is provided by buoyancy, in turn induced by sensible surface heat
fluxes. However given that RICO represents shallow convection over the ocean
(making the Bowen ratio small), the buoyancy becomes dominated by water vapour,
and above $z\approx \SI{200}{\meter}$, the correlation with potential temperature
becomes negative. This transition causes the correlation length-scale to
increase with height until $z\approx \SI{200}{\meter}$, above which the
larger-scale negative correlation takes over. In simulations with ambient shear,
the correlation between vertical velocity and temperature is not only elongated
in the direction of shear, but the correlation is asymmetric in the direction
of shear (evidenced by the skewness and offset of the correlation).
This means that potential temperature features are displaced in the
downwind direction relative to the vertical velocity, suggesting that the
similarity solutions that assume radially symmetric distributions of 
different scalar fields (i.e. all scalar fields are assumed to be a 
function of a single radius $r$), as most plume-based models do
\citep[e.g.][]{Devenish2010}, may not be valid in conditions where shear is present.

\begin{figure*}
\centering
\includegraphics[width=\textwidth]{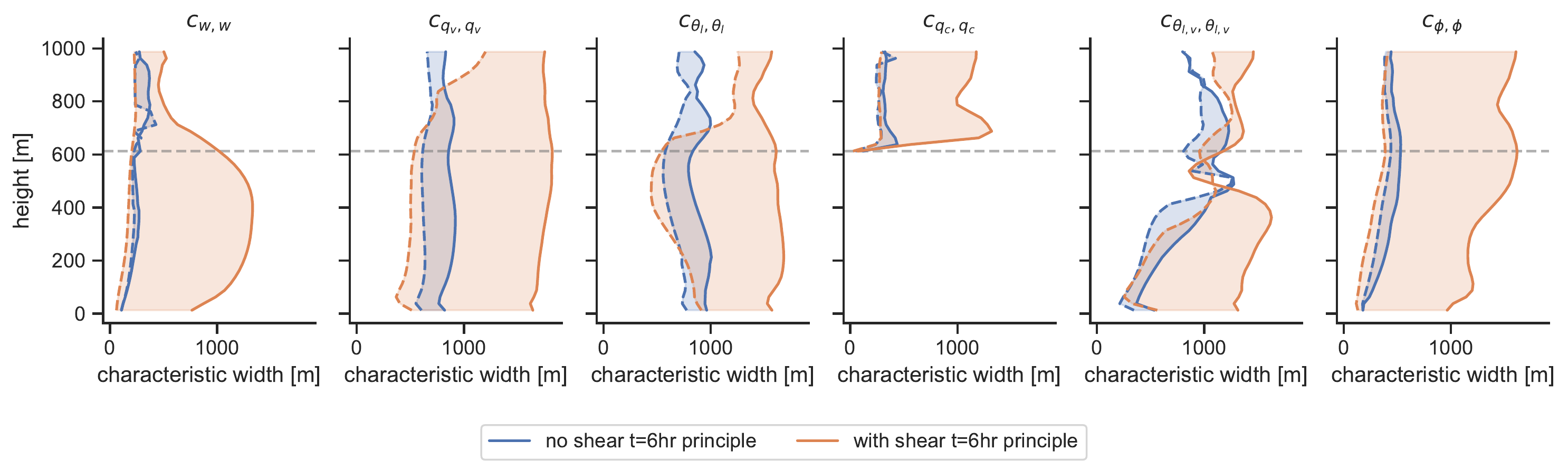}
\caption{\label{fig:cumulant-lengthscale-primitive-fields} Horizontal
coherence length-scales of vertical velocity ($w$), water
vapour ($q_v$), liquid water potential temperature ($\theta_l$), cloud water ($q_c$), virtual liquid water potential temperature ($\theta_{l,v}$) and the radiative passive tracer ($\phi$), for simulations with shear (solid lines) and
without shear (dashed lines).
}
\end{figure*}

To examine the variation in correlation length-scale with height, the
cumulant-based coherence calculation was carried out
at every model level in the boundary layer.
Fig.
\ref{fig:cumulant-lengthscale-primitive-fields} shows the autocorrelation
length-scale as a function of height for both the sheared and non-sheared
simulations. In the discussion below, subscript $S$ and $NS$ will be used to
denote properties extracted from the sheared and non-sheared simulations respectively.
Comparing first the characteristic length-scales across different
fields, the vertical velocity is consistently confined to
narrower features, whereas moisture and sensible heat organise on larger scales \citep[consistent with][]{DeRoode2004}.
The characteristic length-scale of the radioactive tracer field
is generally larger than that for vertical velocity, possibly owing to the tracer
concentration retaining high values in over-turning vortices of thermals within
which the vertical velocity becomes negative.

In both simulations, with and without shear (though more prominent with shear),
the vertical velocity length-scale
increases with height from the surface as thermals are accelerated by the
buoyancy provided through surface fluxes, reaching a maximum scale of
$L_{w,w}^{p}\big\rvert_{NS}\approx \SI{300}{\meter}$ and 
$L_{w,w}^p\big\rvert_{S}\approx \SI{1300}{\meter}$ at $z\approx
\SI{400}{\meter}$. Above this height, velocity scale stagnates (no shear) or decreases
(with shear) with height, as the buoyancy of the rising coherent structures 
becomes negative and the structures begin decelerating through the CIN-layer.
This effect of buoyancy on the correlation
length-scale is more clearly seen when investigating the cross-correlations
with vertical velocity (see below). Comparing the moisture and virtual liquid
potential temperature (buoyancy) length-scales with 
the radioactive tracer, it is notable that although the buoyancy is water-vapour dominated, the length-scales of
correlation of water vapour 
($L_{q_v,q_v}^{p} \big\rvert_{NS} \approx \SI{600}{\meter}$ and
$L_{q_v,q_v}^{p} \big\rvert_{S} \approx \SI{500}{\meter}$ 
at $z\approx \SI{300}{\meter}$)
are significantly larger than those of both the radioactive tracer
($L_{\phi,\phi}^{p} \big\rvert_{NS} \approx \SI{400}{\meter}$ and
$L_{\phi,\phi}^{p} \big\rvert_{S} \approx \SI{300}{\meter}$
at $z\approx \SI{300}{\meter}$) 
and than the
buoyancy features 
($L_{\theta_{l,v},\theta_{l,v}}^{p} \big\rvert_{NS} \approx \SI{500}{\meter}$ and
$L_{\theta_{l,v},\theta_{l,v}}^{p} \big\rvert_{S} \approx \SI{500}{\meter}$
at $z\approx \SI{300}{\meter}$) 
in the lower-half of the boundary layer (until $z\approx \SI{300}{\meter}$).
This means that larger-scale water vapour variability is producing buoyancy on a shorter
length-scale, which in
turn is accelerating boundary layer air on an even shorter length-scale.
This is important
for modelling, as the variability of the buoyancy scalar (here water vapour)
cannot simply be used to infer the scale of coherent rising structures. Lastly, the
maximum correlation length-scale of cloud water (a measure of cloud-size) is on
the order of 
$L_{q_c,q_c}^{p} \big\rvert_{NS} \approx \SI{400}{\meter}$ and
$L_{q_c,q_c}^{p} \big\rvert_{S} \approx \SI{300}{\meter}$
at $z\approx \SI{650}{\meter}$ (near cloud-base), showing most
similarity to the radioactive tracer coherence length-scale at cloud-base.

\begin{figure*}
\centering
\includegraphics[width=\textwidth]{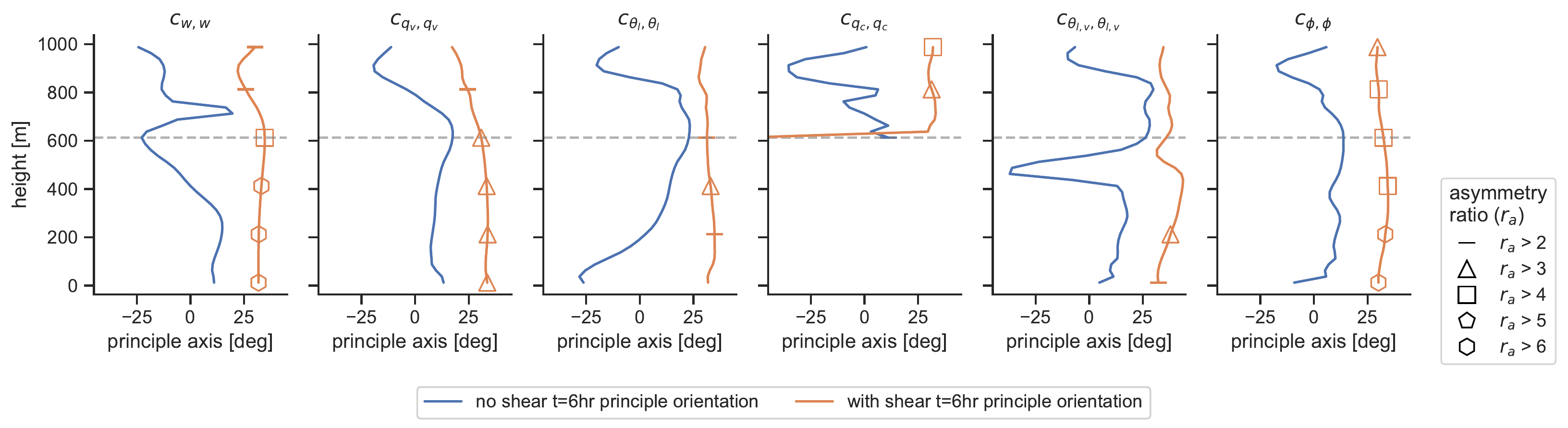}
\caption{\label{fig:cumulant-angles-primitive-fields}
Direction of longest coherence as angle from x-axis (for the variables shown in Fig. 
\ref{fig:cumulant-lengthscale-primitive-fields}) with the degree of asymmetry
given by the marker symbol. In the presence of ambient shear all 
fields are elongated in the direction of shear. In particular, note that cloud structures 
(forming at $z\approx650\si{m}$ are elongated in the same direction as boundary layer structures.}
\end{figure*}

The direction of longest correlation distance (see Fig.
\ref{fig:cumulant-angles-primitive-fields}) is for all fields (including cloud water)
oriented with the ambient wind-direction, which is also the principle direction
of shear, throughout the boundary layer and into cloud-base.
In the absence of shear, some fields
demonstrate some degree of turning with height, but the low degree of asymmetry 
($r_a < 2$ at all heights) suggests that there is very little direction to the
correlation and on examining multiple timesteps the direction of correlation was
seen to be a transient feature, so that no preferential orientation can be
discerned.

\begin{figure*}[p]
\centering
\includegraphics[width=0.85\textwidth]{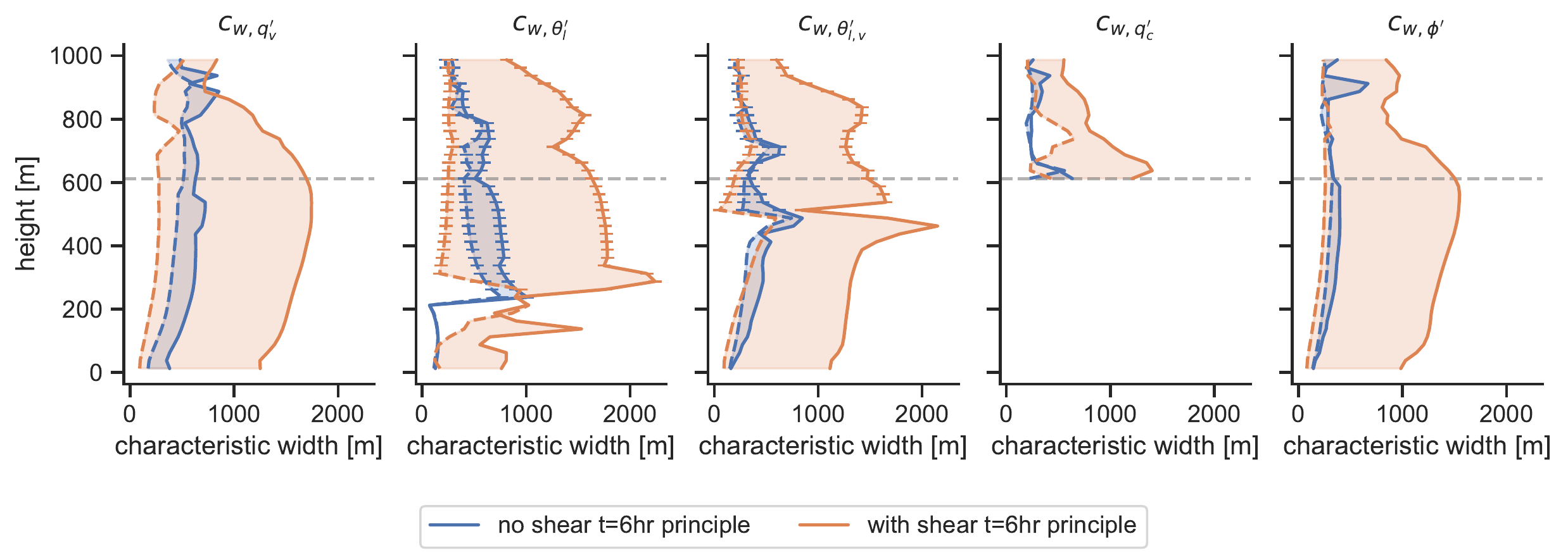}
\caption{\label{fig:cumulant-lengthscales-flux-fields} Horizontal
coherence length-scales, in simulations with (orange lines) and without (blue lines) 
shear, of vertical velocity with water vapour, liquid water
potential temperature, buoyancy (virtual liquid water potential temperature), cloud
condensate and radioactive tracer effectively producing a coherence
length-scale for the vertical flux for each scalar field. Heights at which
anti-correlation occurs are marked with a minus-sign ("-"). 
}
\end{figure*}

Applying the same analysis to the vertical fluxes of heat, moisture, buoyancy
and radioactive tracer (Fig. \ref{fig:cumulant-lengthscales-flux-fields}) we
see that these features are generally narrower than the scalar being transported,
suggesting that the scale of vertical velocity is dominating the length-scales
of vertical transport. All flux fields show clear elongation by ambient shear.
The heights at which the correlation is negative is marked by minus-sign
markers ("-") showing again the transitions for heat (at $z \approx \SI{200}{\meter}$
the heat flux changes sign) and buoyancy (at $z \approx \SI{500}{\meter}$
rising structures are no longer buoyant) where the correlation-length for both
collapses. The moisture and radioactive tracer flux
correlations ($c_{w,q_v}$ and $c_{w,\phi}$) have near-monotonic increases
in size with height until cloud-base is approached.

\subsection{Cross-correlation of scalar fields}
\label{section:cross-correlation-of-scalar-fields}

\begin{figure*}[p]
\centering
\includegraphics[width=0.87\textwidth, right=0.87\textwidth]{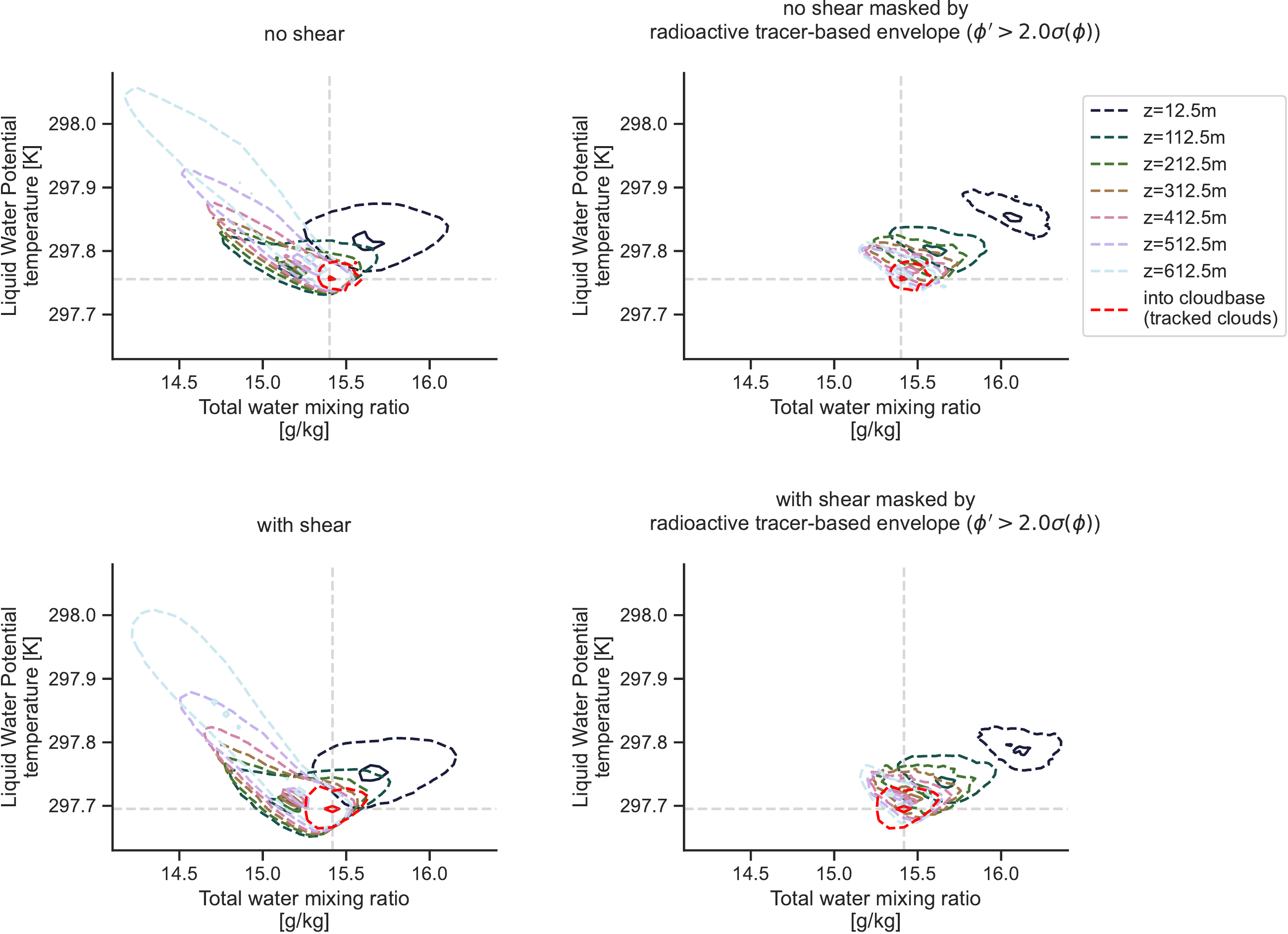}
\caption{\label{fig:cross-correlation-q-t}
Joint distributions of total water vapour specific humidity and liquid potential 
temperature (same as potential temperature in the sub-saturated boundary layer below cloud) in
horizontal cross-sections at increasing heights in the boundary layer, together
with properties of air entering through cloud-base of newly formed 
($t_{age} < \SI{3}{\minute}$) clouds in simulation with (bottom) and without (top) shear, both without
(left) and with (right) the radioactive tracer mask applied. For each height
and at cloud-base, the inner and outermost contours identify the regions
which cumulatively contain the $\SI{10}{\percent}$ and $\SI{90}{\percent}$
highest point density, with the 
number of bins in $\theta_l$ and $q_t$ scaling by the number
of points, $N_{bins}=(N_x \times N_y)^{1/4}$. Grey slashed lines are 
centred on peak of
distributions for cloud-base entering air.
From a relatively moist and warm near-surface state the distribution first dries and
cools into the bulk of the boundary layer, then warms and dries with
height as air subsiding through the boundary layer top imparts a stronger
influence.}
\end{figure*}

As shown above, different scalar fields show
different length-scales of coherence and we now turn to investigating
the extent to which these fields are correlated, not in space but in
the distributions of their scalar values.
In Fig. \ref{fig:cross-correlation-q-t} for the simulations with (right)
and without shear (left), the joint distribution of water vapour
and potential temperature through horizontal cross-sections at
increasing heights are plotted. At each height a bivariate histogram was
created, the bins ranked by number of
points and contours drawn around the bins with 10\% (inner contour, solid
lines) and 90\% (outer contour, stippled lines) cumulative point count.
Constructed in this way the set of contours at each height identify
the centre and width of the joint distribution for the scalar fields
visualised (forming a 2D \textit{box-and-whisker} plot).
For reference, the boundary layer mean values for potential
temperature and moisture are included. In addition to the distributions in
horizontal cross-sections, the joint distribution for points immediately below 
(grid-spacing $\Delta z= \SI{25}{\meter}$) newly-formed ($t_{age} < \SI{3}{min}$) 
clouds (identified by cloud-tracking) is included.

The distributions have similar characteristics in the two simulations; near the
surface the boundary layer is warm and moist relative to the boundary layer
mean, through the super-adiabatic near-surface the distribution rapidly 
becomes cooler and drier with height (until $z \approx \SI{100}{\meter}$).
With further increase in height, the
distribution is stretched to drier and warmer values as mixing with warm dry
air subsiding through the boundary layer top has an increasing influence.

In both simulations, the distribution of air entering newly formed clouds (in
red) coincides with the coldest and most humid part of the boundary layer joint
distribution. The cloud-base distribution contains a larger range of moisture 
values in the simulation with shear, including drier parts of the bulk of the 
boundary layer distribution. This may be because ambient shear increases mixing into the
rising coherent structures, carrying drier air in the regions with high
vertical momentum or because the lower overall virtual potential temperature (buoyancy) 
of the boundary layer causes more air to be buoyant enough to rise to the level of
condensation.

In addition, the distributions in the two simulations differ through a 
translation in
potential temperature by approximately $\approx\SI{0.1}{K}$, which although small
is on the order of the width of the distributions in both cases.
This suggests that in the presence of shear there is a stronger transport
of heat out of the boundary layer and into the cloud layer. 
This offset indicates that even in conditions where the prescribed surface fluxes
are exactly the same, the dynamics of boundary layer transport can vary to a degree
that alters the property of air that forms clouds, and so thresholds based solely
on the thermodynamic
fields (e.g. $\theta_l$ and $q_t$) are inadequate in identifying air that will form 
clouds.

It is clear from these joint distributions that if the aim is to
characterise coherent structures which actually cause the formation of
clouds, it is inadequate to simply construct a conditional sampling based
on threshold values of the scalar fields causing transport. Instead a
method that tracks air transported from the surface layer and into cloud
is required and to achieve this, in the next section, we employ a 
radioactively decaying tracer.

\subsection{Identifying cloud-feeding coherent
structures}\label{identifying-cloud-feeding-coherent-structures}

\begin{figure*}[!b]
\centering
\includegraphics[width=1.0\textwidth]{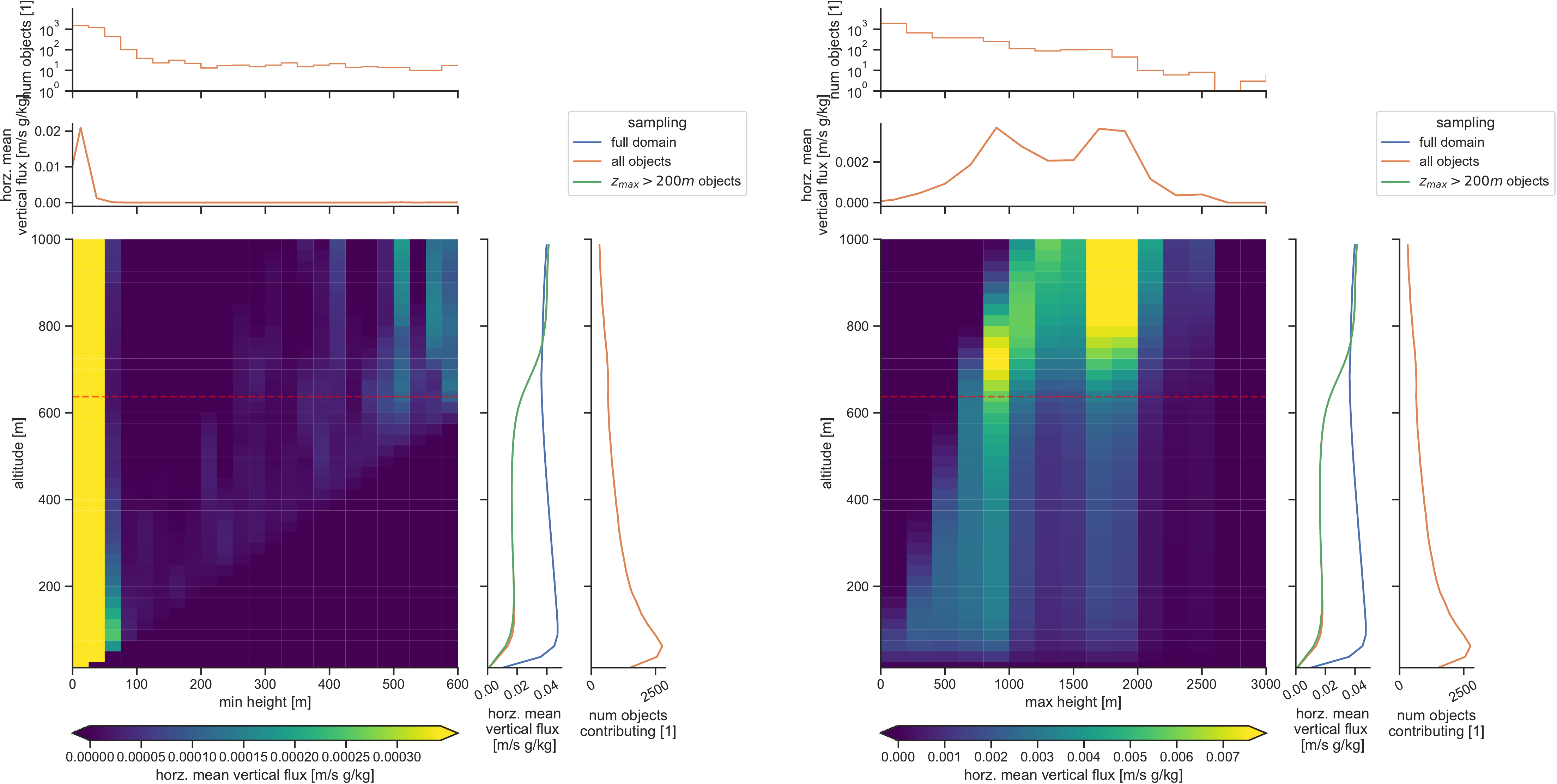}
\caption{\label{fig:qv_flux_decomp_altitude}
Vertical moisture flux decomposed by height and object vertical extent,
decompositing by minimum height (left) and maximum height (right) of each
object in the simulation without shear, with mean flux vertical 
profiles given as right marginal plots, and
boundary layer mean flux and object number distribution as top marginal
plots.
In the moisture flux profile the total domain mean (orange) is shown
together with the moisture flux contributed by the selected objects (blue).
The flux is carried, in aggregate, primarily by objects extending from the
surface and into the cloud-layer, providing flux throughout the boundary layer.
}
\end{figure*}

\label{sec:identifying-cloud-feeding-structures}

We now turn to characterising the coherent structures that have the potential
to trigger clouds.
This is done by first verifying that the air-mass selected by conditional sampling  
using the decaying passive tracer (see \ref{sec:object-identification} for details)
has thermodynamic properties similar to air entering newly formed clouds.
Secondly, we examine the vertical moisture transport by individual coherent
structures (identified by splitting the conditionally sampled air) as a function
of the vertical extent of each structure· This flux decomposition will show that
for coherent structures identified using the decaying passive tracer, it
is the structures which extend from near the surface, through cloud-base and 
into cloud, which carry the bulk of the vertical moisture transport, suggesting 
that the clouds manifest as billowing tree-crowns sat atop trunks of 
continuous transport.

Examining first the thermodynamic properties of air sampled using the decaying
passive tracer, the joint distributions of the previous section are
conditionally sampled (Fig. \ref{fig:cross-correlation-q-t} right), by requiring
that the concentration of decaying passive tracer is at least 2 standard
deviations from the horizontal mean (see Sec.
\ref{sec:object-identification} for details). 
Using this method for both the sheared and non-sheared simulations,
the distributions collapse down to align near-perfectly with that of air
entering through cloud-base, indicating that the radioactive tracer is picking
exactly the air that may trigger clouds.  
In addition, for both cases the means
of the distributions appear to be near-linearly translated with height
suggesting that these coherent cloud-triggering structures mix with the bulk of
the boundary layer at a similar rate for both potential temperature and
water vapour.
Finally, the widths of
the joint distributions appear nearly unchanged with height. All of these
facts are encouraging for the prospect of parameterisation of the mixing into
coherent updrafts in the boundary layer, by which the properties of
cloud-triggering air may be predicted based on the surface fluxes (and other
external forcing factors).

Although the radioactive tracer method identifies air with the same
statistical properties as air that is triggering clouds, it is not
guaranteed that every volume of this boundary layer air will actually
trigger a cloud. Some structures may simply be too small to survive the
journey to the condensation level without being completely mixed into
the bulk boundary layer air. This can be addressed in further work by
tracking boundary layer structures and identifying which ones actually
trigger clouds (beyond the scope of this study). Here, we instead 
identify the structures that dominate the vertical
moisture transport.

We examine this next by decomposing the moisture flux at a given height by the
vertical extent (measured as the height of the bottom $z_{min}$ and the top $z_{max}$)
of coherent structures contributing to the flux in \autoref{fig:qv_flux_decomp_altitude}. Both the simulations with and without shear 
show the same structure, and so we only show the situation without shear here.
Note that an individual object of a given vertical extent, will likely contribute
to the flux at a number of different heights and the figure simply shows
at any given height how high (\autoref{fig:flux_decomp_by_wt_and_fp} right)
or low (\autoref{fig:qv_flux_decomp_altitude} left) structures which
contribute to the flux extend.

Considering first the flux decomposition by the minimum height of each
coherent structure (\autoref{fig:qv_flux_decomp_altitude} left)
we find that almost all flux is carried by structures
extending down to the surface ($z \approx \si{100}{m}$, even though this
accounts for only half of the number of coherent structures.
Considering secondly the maximum height of the coherent structures we
find that $\approx 90\%$ of the moisture flux through the boundary
layer is transported by structures which extend above cloud-base ($z_{cb} \approx \si{650}{m}$),
transporting moisture all the way from near the surface into and
through cloud-base. The flux is dominated by structures terminating at
$z \approx \si{900}{m}$ with a second peak at $z \approx \si{1800}{m}$,
possibly accounting for clouds reaching and passing the height of free
convection \citep[\textit{forced} and \textit{active} clouds as denoted by][]{Stull1985}.
From the above analysis we conclude that were we to characterise
coherent structures solely by the passive tracer we would be
characterising in part the properties of structures doing transport in
cloud. As we are concerned here primarily with the characteristics of
structures that effect transport below cloud, we will in later sections
crop structures at cloud-base so that only the sub-cloud part is
considered.

\subsection{Minkowski characteristics of coherent
structures}\label{minkowski-characteristics-of-coherent-structures}

Having identified the below-cloud coherent structures with correct thermodynamic
properties (for triggering clouds) and carrying the majority of the vertical
moisture, we next calculate characteristic properties of each
structure and decompose the boundary layer moisture flux by the object
scales, to ascertain the characteristics of structures dominating the flux.
This flux decomposition is visualised as a (boundary layer mean) moisture-flux
weighted probability density distribution (PDF)
\citep[computed using Gaussian Kernel Density Estimation][]{Jones2001}, with
weights given by the per-object total moisture-flux and the PDF computed
over the object characteristics used in the decomposition.
To demonstrate the effect of external forcing the characteristics of structures
dominating the flux in the simulations with and without shear will be
contrasted.
To do this we calculate a characteristic 
length, width and thickness for each object using the so-called Minkowski
functionals (detailed in Sec. \ref{sec:methods-minkowski-functionals}).

\begin{figure}
\centering
\includegraphics[width=1.0\textwidth]{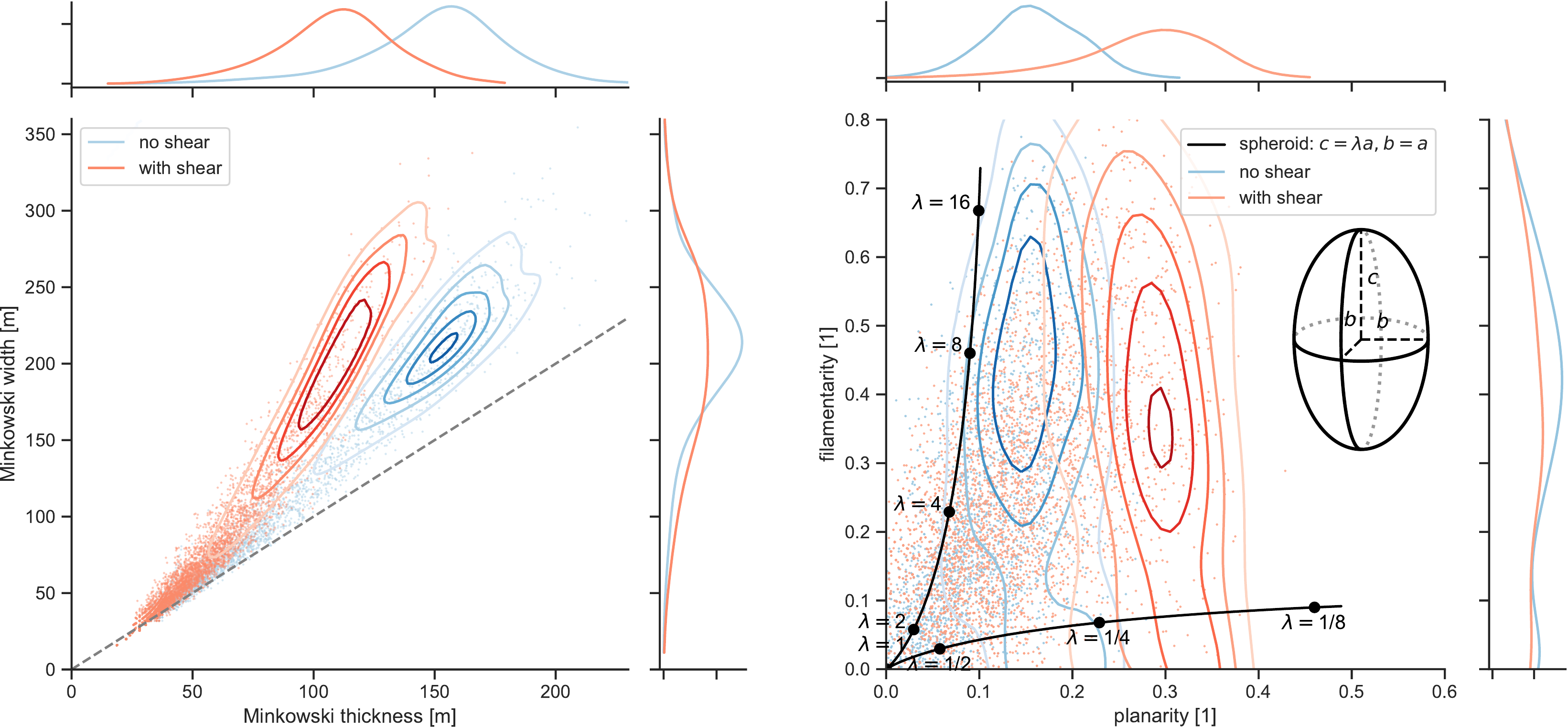}
\caption{
Minkowski functional derived width vs thickness (left) and filamentarity vs planarity (right)
of individual objects (points) for the cases with (red) and without (blue) shear.
The contours depict the distribution of vertical moisture flux with these object
characteristics (computed as a flux-weighted probability density functioned
constructed using Gaussian Kernel Density Estimation). In the filamentarity vs planarity
plot a spheroid is given as reference reference (solid black) with independent axis
parameterised by $\lambda$ and deformation of a spheroid through a general ellipsoid with
fixed major axis and varying aspect between the remaining two axes ($\alpha$).
\label{fig:flux_decomp_by_wt_and_fp}
}
\end{figure}

We first examine the boundary layer moisture flux when decomposed by the
width and thickness of the coherent structures (Figure \ref{fig:flux_decomp_by_wt_and_fp} left)
with contours in red and blue depicting the distribution in the cases
with and without shear respectively. The scatter-plot, depicting the width and
thickness of each structure, shows that the majority of
structures are small and nearly axially symmetric (appear near the
unit line). However, once we consider the contribution to the vertical
moisture flux the larger and more asymmetric structures dominate. This
asymmetry is more pronounced when ambient shear is present, shifting
the width/thickness ratio from $\approx \si{150}{m}/\si{200}{m} = 3:4$
to $\approx \si{100}{m}/\si{200}{m} = 1:2$.

This asymmetry can be succinctly captured by computing the filamentarity
and planarity (measuring how \emph{pencil}-like or \emph{disc}-like
object each is) as seen in \autoref{fig:flux_decomp_by_wt_and_fp} right, showing
that ambient shear causes the coherent structures to be stretched
planar, which can be seen by a $\approx \SI{100}{\percent}$ 
increase in planarity from $P_{NS}\approx 0.15$ and $P_{S} = 0.35$. 
As a reference (in black), the filamentarity and planarity of an ellipsoid with 
varying elongation (parameterised as the aspect ratio between one axis and the 
remaining two) the coherent structures can be seen to move from being more
cylindrical to sheet-like in shape. Here only structures with Euler characteristic
$\chi = 1$ (objects without holes) are considered (as the length is otherwise
not defined). Although objects with holes do contribute a significant fraction
of the moisture flux ($\approx 20\%$ and $\approx 70\%$ for the cases without 
and with shear respectively) this fraction is reduced when the passive tracer
threshold is increased (to $\approx 4\%$ and $\approx 30\%$ for $\phi=3$) 
and the relative stretching between the two cases becomes more pronounced (not shown here).

\subsection{Object orientation}

\begin{figure}
\centering
\includegraphics[width=0.5\textwidth]{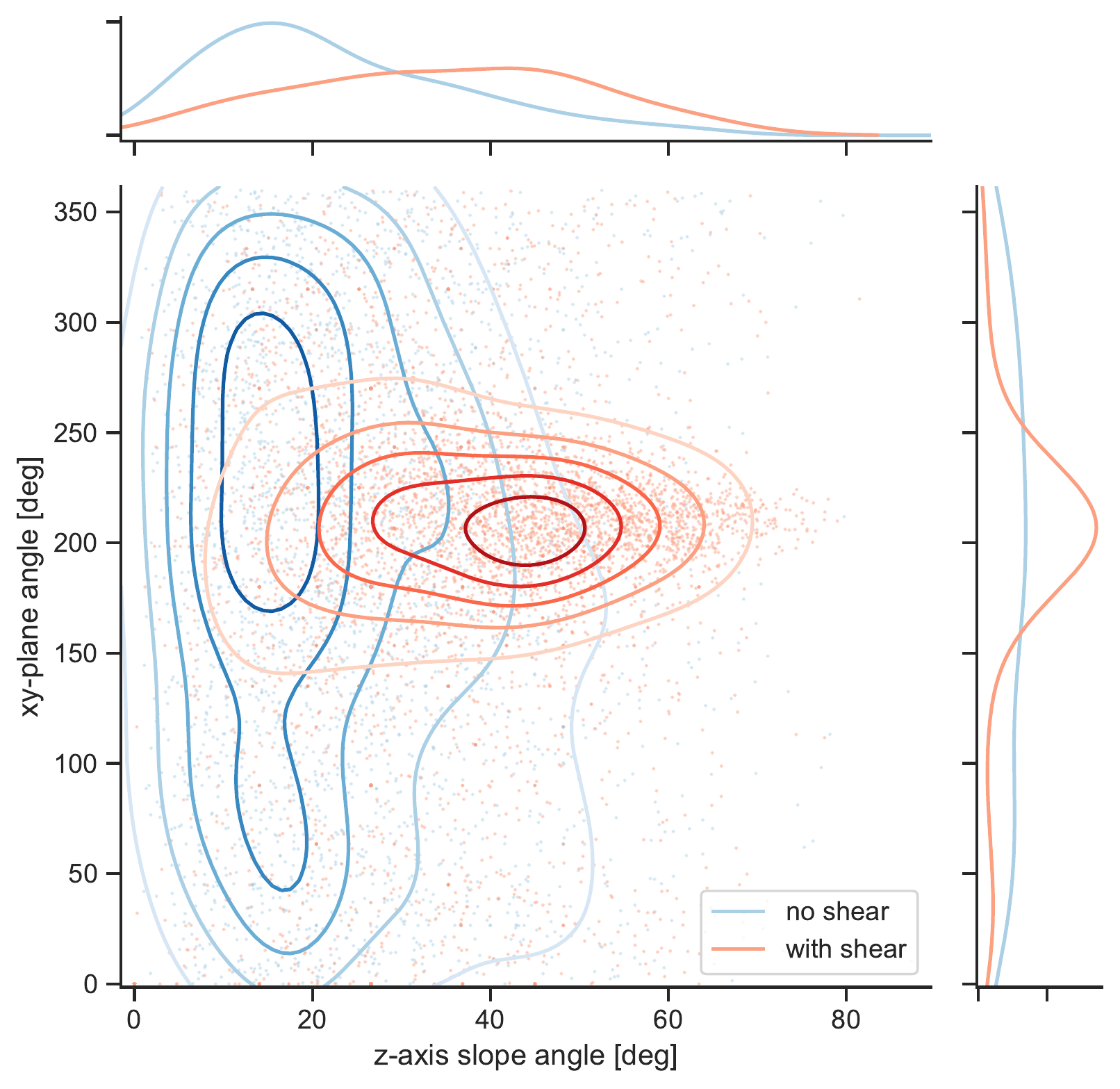}
\caption{
Vertical moisture flux decomposed (as in \autoref{fig:flux_decomp_by_wt_and_fp})
by xy-orientation angle ($\phi$) against z-axis 
slope angle ($\theta$) of coherent structures
present at $t = \SI{6}{\hour}$ in simulations with (red) and without shear (blue), together
with distributions in each along plot margins
\label{fig:object-properties-orientation-and-tilt}
}
\end{figure}

In addition to knowing the characteristic length-scales and shape of
individual coherent structures, it is instructive to determine the tilt and orientation
of each object to be able to formulate an integral model to represent transport
by coherent structures. In Figure \ref{fig:object-properties-orientation-and-tilt},
the vertical moisture flux has been decomposed by the
tilt and orientation angles of all sub-cloud objects present at $t = \SI{6}{\hour}$ in both
the simulation with and without shear.
This shows the presence of ambient
shear ($\approx \SI{2}{\meter\per\second}$ change in wind-speed over the sub-cloud layer) 
caused the mean tilt of individual objects to change
from $\theta_{no,shear} \approx \SI{15}{\degree}$ 
(near-vertical, given the near symmetrical angular spread
in orientation angle) to $\theta_{shear} \approx \SI{40}{\degree}$ and
changed the structures from having no preferential horizontal
orientation, to being oriented with direction of wind shear
($\phi_{shear} \approx \SI{210}{\degree})$.

This direction of orientation in the presence of ambient wind was found to
coincide with the principle direction of coherence identified using
cumulants (\autoref{sec:results-characteristic-length-scales}).
Being able to quantify the stretching and tilting of individual structures is an
improvement on measure of elongation provided by using cumulants as we
are able to separate out cumulant elongation due to individual structures
from that caused by structures being spatially organised into linear
features.

\section{Discussion}\label{discussion}

\label{sec:discussion}


The two methods presented here, the first focusing on bulk-measures of coherence
in the boundary layer (using cumulants) and the second on identifying and
quantifying morphological properties of flux dominating individual 
coherent structures (using Minkowski functionals and tilt/orientation calculation), 
have complementary strengths.

The principal difference
between the two approaches is that the cumulant-based method produces a
length-scale estimate through considering the spatial coherence throughout the
fluid, whereas identifying and characterising individual objects gives an
estimate of scale for individual coherent structures. Two individual objects in close
proximity will increase the cumulant length-scale estimate, meaning that the
cumulant length-scale conflates the object size and spatial organisation
of objects (specifically the inter-object distance).  
In addition, in cases where 
there are multiple populations of coherent structures, the cumulant 
method will conflate these into one integrated measure, 
and the relevant details of each separate population may be lost. In 
particular, in the upper part of the boundary layer which is characterised 
by ascending buoyant thermals and descending entrainment flows, the cumulant 
analysis will not necessarily give a "clean" description of either of these.

Another aspect by which the two approaches differ is how the cumulant method
estimates length-scales in the horizontal plane (the method requires
translational symmetry to study coherence as a function of displacement
and so cannot be applied vertically without picking a reference height), whereas the
Minkowski length-scales of individual objects are not constrained in the orientation in
which these length-scales are calculated. This means that the two
measures of length cannot be directly compared, as the Minkowski length-scales
are not measured in the horizontal plane and necessitate the calculation of
object orientation to interpret the length-scales calculated.
It further means that the cumulant-based measure of scale cannot separate out
the orientation of individual structures from the orientation of spatial
organisation in general. The degree of tilt could though be measured with cumulants
by computing correlation between different heights with distance in the
horizontal plane (not shown here).

Both techniques demonstrate that the spatial characteristics of coherent boundary
layer structures are affected by the ambient wind. With the cumulant-based technique
this manifests as an elongation across all scalar fields in the direction of shear.
Through decomposition of the vertical moisture flux by the vertical extent
of each structure, we showed that the majority of the flux is carried by structures
which extend from the near-surface and into the cloud layer. This suggests that
these structures are more plume-like (providing near-continuous flux as a function
of height) rather than thermal-like. The ambient wind causes these flux-dominating
coherent structures from being near axisymmetric to being stretched planar and 
tilted in the direction of ambient wind.

These findings suggest that models representing non-local transport by coherent boundary
layer structures, should choose the plume model as the fundamental starting point.
The effect of ambient shear appears to be to organise vertical transport into coherent 
structures which are tilted and stretched
planar (both of which increase entrainment by increasing the surface area),
and at the same time organised into linear features (which may 
decrease entrainment by limiting the dry air reaching individual plumes).

\section{Conclusions and further work}
\label{conclusions-and-further-work}

This paper has demonstrated two methods by which to characterise
the shape, size and orientation of coherent structures.
The first method quantifies the horizontal orientation and length-scale of coherence
between any two scalar fields, and through this makes it possible to
measure the coherence in the boundary layer as a whole. The second
method identifies cloud-feeding coherent structures using a decaying
passive tracer and is able to quantify length-scales and orientation
(both vertical and horizontal) for each of these objects, allowing for
a more instructive decomposition where the non-local transport by
individual coherent structures can be studied.


We have demonstrated the use of cumulants to measure characteristic
length-scales for different scalar fields and fluxes of these fields in
the bulk of the boundary layer. This showed that in the absence of shear,
vertical velocity features were significantly narrower ($\approx\SI{200}{m}$)
than the moisture and potential temperature fields ($\approx\SI{1000}{m}$), but
of a similar scale to a surface-released passive tracer ($\approx\SI{300}{m}$).
This method is also able to quantify the elongation of spatial coherence and
calculate the elongation direction, showing how the presence of
ambient wind shear causes elongation (along with the wind-direction in
this case) of the vertical velocity field and
a less-pronounced change to the heat and moisture fields.

We additionally demonstrated how a surface-released radioactive tracer may be
used to identify air with thermodynamic properties statistically similar
to air entering through cloud-base of newly-formed clouds.
This allowed for the identification of individual cloud-feeding coherent
structures. By decomposing the vertical flux by the vertical extent of
each structure it was found that the vertical flux is dominated by structures
extending from the surface, through cloud-base and into cloud. This insight
suggests that flux-dominating structures are more plume-like (carrying
transport from the surface and throughout the boundary layer) rather than
thermal-like. The part of these structures carrying out transport in the
boundary layer were then characterised using Minkowski functionals 
(producing a characteristic length, width and
thickness for each object) and a technique for calculating an object's
tilt and horizontal orientation.
With these methods it was showed that although the majority of coherent structures
are closest to a sphere in shape, the structures which dominate the vertical
flux in the boundary layer are generally thicker than they are wide ($4:3$ ratio);
an asymmetry which increases in the presence of ambient shear ($2:1$) as
the structures are stretched planar. In absence of shear the structures do
generally exhibit some degree of tilt ($\theta \approx \SI{10}{\degree}$), but 
this has no particular orientation, whereas in the presence of shear objects
were more tilted ($\theta \approx \SI{40}{\degree}$) and all in the direction
of the ambient wind.
These findings cannot be made through
a bulk-estimate of length-scale of coherence, as was done in previous
work, but requires identifying and measuring individual coherent structures.


These findings suggest that when formulating models to represent transport
by cloud-feeding coherent boundary layer structures, the most appropriate
model may be a plume, at least for the kind of (shallow moist convective) 
boundary layer studied here. 
In the presence of shear, the planar stretching and tilting
\citep[possibly increasing the entrainment of dry air]{Bursik2001} of these structures 
may be necessary to consider when constructing a parameterisation of boundary layer transport.
Further study can explore the generality 
of this conclusion to other boundary layer configurations, e.g. land-based, 
heterogeneously forced, deep convective etc. Similarly the techniques presented
herein may be used to study the relationship between the spatial characteristics
of coherent boundary layer structures and the spatial characteristics of the 
clouds they form.

This work used the ARCHER UK National Supercomputing Service (www.archer.ac.uk) 
and was funded through the NERC/Met Office Joint Programme "Understanding and Representing Atmospheric Convection across Scales (ParaCon)", grant NE/N013840/1. 
The software implementation of
the techniques presented here are available in \citet{Denby2020genesiscode}.

\appendix

\section{Time-scale of convective overturning in boundary layer}
\label{appendix:convective-timescale}

Using the sub-cloud characteristic velocity scale $w_{*}$ 
\citep[as in][but corrected for the contribution to buoyancy from water vapour]{HoltslagNieuwstadt1986}
and the boundary layer depth $z_{BL}$ we can calculate a
sub-cloud convective overturning time-scale $\tau_{BL}$ as

\begin{equation}
    \tau_{BL} = \frac{z_{BL}}{w_*}
\end{equation}

The sub-cloud convective velocity scale is given as

\begin{equation}
    w_* = \left( \left.\overline{w'b'}\right|_0 z_{BL} \right)^{1/3},
\end{equation}
with buoyancy flux

\begin{equation}
\left.\overline{w'b'}\right|_0 = \frac{g}{T_{v,0}} \left.\overline{w'\theta_v'}\right|_0,
\end{equation}
and virtual potential temperature flux (approximately) given by \citep{DeRoode2004}

\begin{equation}
    \left.\overline{w'\theta_v'}\right|_0 = \left.\overline{w'\theta'}\right|_0 + \frac{R_v}{R_d} \overline{\theta}_0 \left.\overline{w'q'}\right|_0,    
\end{equation}
where $z_{BL}$ is the boundary layer depth, $g$ the gravitational acceleration 
and $|_0$ denotes surface values. With surface fluxes for sensible
$\left.\overline{w'\theta'}\right|_0 = \frac{F_s}{\rho_0 c_{p,d}}$ 
and latent heat $\overline{\theta}_0 \left.\overline{w'q'}\right|_0 = \frac{F_v}{\rho_0 L_v}$,
surface moisture $q_{v,0} = \SI{15}{\gram\per\kilogram}$,
surface temperature $T_0 = \SI{300}{\kelvin}$ and 
diagnosed boundary layer depth $z_{BL} = \SI{650}{\meter}$ (here taken as the 
cloud-base height) the convective overturning time-scale becomes $\tau_{BL} = \SI{16}{min}$. 
The constants used above are those for density of dry air 
$\rho_0 = \SI{1.2}{\kilogram\per\cubed\meter}$, 
latent heat of vaporisation $L_v = \SI{2.5e6}{\joule\per\kg}$, 
specific heat capacity of dry air $c_{p,d} = \SI{1005}{\joule\per\kilogram\per\kelvin}$, 
gas constants for dry air $R_d = 287 \SI{287}{\joule\per\kilogram\per\kelvin}$ 
and water vapour $R_v = \SI{461}{\joule\per\kilogram\per\kelvin}$.

\section{Crofton's formula for discrete integrals}
\label{appendix:croftons-formula}

Numerically evaluating the integrals of the Minkowski functionals in 3D (equations \ref{eqn:mink_0} to \ref{eqn:mink_3}) is non-trivial on discrete 3D masks as these structures are necessarily blocky and so constructing, for example, the surface normal at a corner is poorly defined.

Instead of approximating the surface normals, the integral in the Minkowski functionals can be approximated discretely using Crofton's formula \citep{Schmalzing1999}. This amounts to counting the number of vertices ($N_0$), edges ($N_1$), faces ($N_2$) and cells ($N_3$) on both the interior and exterior of a given object mask. With these the Minkowski functionals in 3D are given as

\begin{align*}
    V_0 &= N_3,
    \ V_1 = 2 \frac{N_2 - 3 N_3}{9 \Delta x},
    \ V_2 = 2 \frac{N_1 - 2 N_2 + 3 N_3}{9 \Delta x^2}, \\
    V_3 &= \frac{N_0 - N_1 + N_2 - N_3}{\Delta x^3}.
\end{align*}

It can be shown for shapes where analytical forms for the Minkowski functionals exist that these approximate definitions converge to the true values when $\Delta x \to \infty$. \textbf{Note} that the above approximations assume the underlying grid to have isotropic grid spacing ($\Delta x = \Delta y = \Delta z$).

\section{Calculation of slope and orientation of individual objects}
\label{appendix:slope-and-orientation-calc}

The $xy$-orientation angle ($\phi$ measured from the $x$-axis) and tilt angle $\theta$ (measured from the $z$-axis) are calculated from characteristic slope scales $\overline{\Delta x}, \overline{\Delta y}, \overline{\Delta z}$:

\begin{align*}
    \phi = \arctan2(\overline{\Delta y}, \overline{\Delta x}), \
    \theta = \arctan2(\overline{\Delta l_{xy}}, \overline{\Delta z}), \
    \Delta l_{xy} = \sqrt{\overline{\Delta x}^2 + \overline{\Delta y}^2},
\end{align*}
which are evaluated as area-weighted changes in the centroid position

\begin{align*}
    \overline{\Delta x} = \frac{\Delta z}{V} \sum_{k} A(z_k)\frac{x^c(z_{k+1}) - x^c(z_{k-1})}{2}, \
    \overline{\Delta y} = \frac{\Delta z}{V} \sum_{k} A(z_k)\frac{y^c(z_{k+1}) - y^c(z_{k-1})}{2}, \
    \overline{\Delta z} = \frac{\Delta z}{V} \sum_{k} A(z_k)\frac{z^c(z_{k+1}) - z^c(z_{k-1})}{2},
\end{align*}
where $x_c(k)$, $y_c(k)$ and $A(k)$ are the centroid x and y position, and
area at height-index $k$ respectively, and $V$ the volume, given by
\begin{align*}
    A(z_k) = \sum_{i, j} m(x_i, y_j, z_k) \Delta x \Delta y, \ V = \sum_{i,j,k} m(x_i, y_j, z_k) \Delta x \Delta y \Delta z\\
    x^c(z_k) = \frac{\sum_{i, j} x_i\ m(x_i, y_j, z_k)}{\sum_{i, j} m(x_i, y_j, z_k)}, \ 
    y^c(z_k) = \frac{\sum_{i, j} y_i\ m(x_i, y_j, z_k)}{\sum_{i, j} m(x_i, y_j, z_k)}
\end{align*}
for an individual object defined by the mask $m$ and grid-spacing $\Delta x, \Delta y, \Delta z$.

\addcontentsline{toc}{section}{References}

\bibliography{library.bib}


\end{document}